\documentclass[10pt]{article}
\usepackage{amsmath}
\usepackage{verbatim}
\usepackage[dvips]{graphicx}
\usepackage{amssymb}
\usepackage{amsfonts}
\usepackage{amsthm}
\usepackage{latexsym}
\usepackage{geometry}
\begin{document}
\author{Wenguang Mao \and Xudong Wang}
\title{Scalable Two-way Relaying}
\date{}
\maketitle

\section{Introduction}
Two way relaying is a promising technique to improve the performance of wireless networks because it dramatically enhances the spectral efficiency comparing to traditional one way relaying. This technique is flourished by seminal works \cite{hottopic} \cite{katti} \cite{af} \cite{spectral}, which introduce different two-way relaying schemes. Generally, these schemes can be divided into two categories: three-step schemes and two-step schemes. In three-step scheme also known as decode-and-forward (DF) relay, two end nodes in two way relay channel transmit their packets sequentially in the first two time slots and the relay node codes two received packets by applying bitwise XOR operation and broadcasts the coded packet to two end nodes in next slot. In two-step schemes also known as amplify-and-forward (AF) relay or denoise-and-forward (DNF) relay depending on the operation applied in the relay node, two end node send packets concurrently to the relay node in the first time slot and the relay node directly broadcasts the received waveform (AF relay) or perform physical layer network coding (PLNC) on received signals and broadcast the coded packet (DNF relay) to two end nodes in the second time slot. As shown in \cite{af}, two-step schemes are more efficient than three-step scheme especially in high SNR regime. Hence in this paper we only take two-step two way relaying schemes into consideration.

Motivated by enhancing the performance of the network, many research papers investigate how to apply the two way relaying technique in wireless networks. Most of these papers focus on networks with regular topologies, such as star topology \cite{star}, layered topology \cite{layered}, two-tier topology \cite{BS}, a routing path in multihop networks \cite{linear}, source-relays-source topology \cite{ofmd2,ofdm3} and etc. For general ad hoc networks, literature \cite{multihop} provides a solution for applying two-way relaying in such networks based on transmission scheduling. However the optimal scheduling in this paper turns out to be NP-hard, and similar conclusion is further confirmed in \cite{schedule}. Hence the scheduling based solutions for applying two-way relaying are more than complicated in general ad hoc network, and are not practical in realistic systems. Therefore to apply two way relaying in general networks in a more practical and more scalable way, random access schemes should be taken into consideration.

Recently, a few papers deal with topics related to the design of random access MAC protocol to support two-way relaying. Majid Khabbazian \cite{MACtheory} presents a MAC design for analog network coding (similar with AF relay). However the solution is on theoretical level and far from a practical design. In addition, Shiqiang Wang \cite{DMAC} proposes a distributed MAC protocol to enable the application of PLNC in the network. However, the protocol suffers from some defects. First, the protocol has no mechanism to guarantee the performance gain when there are no bi-direction data flows, namely when not both two end nodes have packets to each other the throughput would degrade to the level when no two way relaying is applied. This limits the usage of this protocol. Also, in the protocol the transmission is initiated by the relay node instead of transmitters of data. This requires that nodes have to provide their queue information to neighbor nodes, and this operation would result in the degradation of the performance due to the overhead and the increase of mean packet delay. To our best knowledge, no better solution schemes are published up to now. Hence a more practical and more efficient random access MAC protocol which supports two way relaying is high needed.

To support random access MAC protocol, physical layer implementation of two way relaying should be revised. Some requirements of two way relaying such as the symbol synchronization and frame alignment are difficult to meet in random access protocol, and hence should be removed. Some previous works \cite{katti} \cite{ofdm} \cite{ConPNLC} \cite{ConANC} \cite{bpplnc} \cite{BCJR} remove or relax these requirements and make the two way relaying technique more practical. However, these solutions have some imperfections respectively and are insufficient to serve as powerful physical layer schemes to support random access MAC protocol. In \cite{katti}, analog network coding scheme is proposed without any synchronization requirement. However, the scheme is designed only for MSK modulation and cannot support QAM modulation which is more widely used. The OFDM based solution, proposed in \cite{ofdm}, only allows the asynchronization of packets up to the length of cyclic prefix of the OFDM symbol. This limits the usage of this method in many scenarios. The schemes based on linear convolution coding \cite{ConPNLC} \cite{ConANC} suffer from the similar issue, i.e. the scheme only can tolerate the symbol level misalignment but frame synchronization is still required. Also asynchronous two way relaying schemes based on PLNC, proposed in \cite{bpplnc} \cite{BCJR}, have the problem that coding scheme is sensitive to channel gains and modulation schemes adopted by two end nodes as revealed in \cite{constellation}. The author of \cite{constellation} also shows that the issue is further complicated when large constellation size is used. This inflexibility results that these schemes are not suitable for random access networks especially mobile ones. Therefore to provide more freedom on the design of a widely applicable random access MAC protocol, it is necessary to design a new physical layer scheme for two way relaying, which is flexible, fully asynchronous and feasible for multiple types of modulation schemes.

In this paper, we propose not only a random access MAC protocol to support two-way relaying but also a practical physical layer schemes for two-way relaying to facilitate the MAC design. To our best knowledge, we are the first to provide a integrated design including both physical layer and MAC layer for applying two way relaying in general ad hoc networks. This integrated design deliveries a practical and high-performance solution. Specifically, we first present a new physical layer decoding algorithm for end nodes in two way relay channel. The algorithm is based on a bunch of techniques such as oversampling, joint channel estimation and waveform recovery. This decoding algorithm in end nodes and the amplify-and-forward operation in relay node compose the our physical layer scheme. The advantage of our physical layer scheme is manifold. First, our scheme does not require any synchronization. Second, our physical layer is feasible for any linear modulation schemes. Also, all of information required for physical layer operations, e.g. channel coefficient, are locally obtainable. Hence our physical layer scheme adds little extra burden on MAC protocol. Then we propose a random access MAC protocol TREAN (Two-way Relaying Enhanced Ad-hoc Network). It performs RTS/CTS-like queries to build the cooperative configuration for two-way relaying. After that the new physical layer scheme is applied to conduct two-way relay transmission with least coordination. Also, different modes of TREAN are designed to enlarge the application scope of TREAN protocol. The basic mode is simple, flexible and requires least management. It works well in networks with bi-directional data flows and is also suitable for mobile network. The extended modes of TREAN protocol is designed to deliver high performance even when bi-directional data flow is absent. Moreover, we provide accurate analysis and approximate derivation on the saturation throughput for small-scale network and large-scale network respectively. These theoretical results can serve as tools for performance evaluation or guidelines for practical designs. To validate our design and analysis, we perform the simulation in various settings. Results show that TREAN protocol can significantly enhance the throughput of the network. Also, the theoretical analysis is verified by the fact that analytic results and simulation results are matched well.

The rest of this paper is organized as follow: Section \uppercase\expandafter{\romannumeral2} introduces our physical layer scheme; Section \uppercase\expandafter{\romannumeral3} describes details of the TREAN protocol; Section \uppercase\expandafter{\romannumeral4} provides throughput analysis when TREAN protocol is adopted in the network; Section \uppercase\expandafter{\romannumeral5} evaluates the performance of our integrated design; Section \uppercase\expandafter{\romannumeral6} concludes the paper.
\section{New Physical Layer Decoding Scheme for two way relaying}
In our physical layer scheme for two way relaying, two end nodes transmit their packets to the relay node without any synchronization. After receiving superposed packets, the relay node simply amplifies and forwards the received signals to end nodes. Then two end nodes extract their desired packets from superposed waveform broadcast by the relay node with our new decoding algorithm. In this section, we present this new decoding algorithm. The basic idea of this new physical layer technique is to exploit the oversampling. First, the superposed waveform broadcast by the relay node is sampled by the receiver with the frequency higher than symbol rate. Then, we deduct the components of the known packet from these samples. Finally, we can recovery the waveform of unknown packet with processed samples according to Shannon sampling theory and decode the obtained waveform with general decoding procedure. The details of our new physical layer scheme are explained as follow.

\begin{figure}[ht]
\centering
\includegraphics[width=0.5\textwidth]{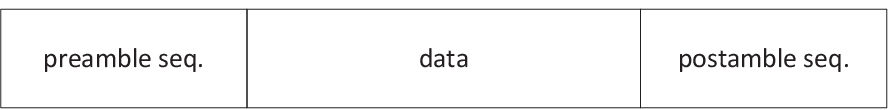}
\caption{\label{preamble} Packet layout for our physical layer technique. The data field consists of PLCP header added in physical layer and the frame from MAC layer.}
\end{figure}

\begin{figure}[ht]
\centering
\includegraphics[width=0.75\textwidth]{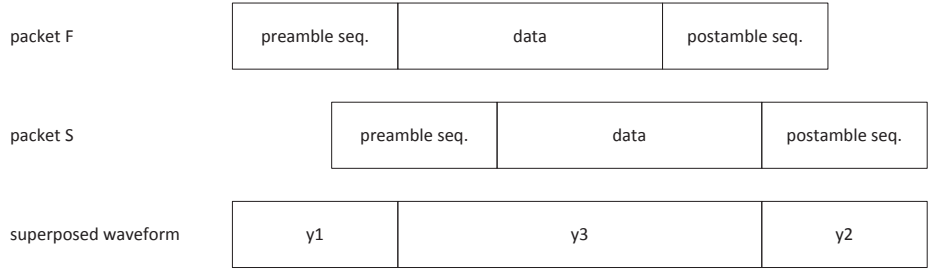}
\caption{\label{jdec} Superposed packets. We can observe that the preamble sequence of packet F is not collided by the data field of packet S, while the postamble of packet S is free from the interference of the data field of packet F. }
\end{figure}

\paragraph{Sampling and packet detection} First of all, we show the packet layout for our physical layer as Figure \ref{preamble}. Similar with the design in \cite{katti}, we divide the training sequence field into two parts and add them at the beginning and the ending of the packet as preamble sequence and postamble sequence respectively. The preamble is a $L_{p}$-bit pilot sequence $\{p_{n} \} \ (1 \leqslant n \leqslant L)$ and the postamble is another one which is orthogonal to the preamble sequence. The objective of this layout is to avoid the situation that the pilot sequence of one packet is completely collided with the data field of another packet when two packets are superposed at the relay node. This situation may result in the failure of channel estimation. However, if two packets have similar sizes\footnote{This can be easily guaranteed with the help of upper layer protocol. For example, in TREAN protocol, the two-way relaying cooperation only happens between frames with same type and hence with similar sizes.}, with our design at least one pilot sequence, preamble sequence or postamble sequence, would be almost free from the interference from the data field of another packet, as shown in Figure \ref{jdec}.

Under random access MAC protocol such as 802.11 and TREAN, the receiver cannot know when a packet arrives. Hence the receiver has to keep sampling the channel to detect the arrival of packets. In our physical layer, we require that the receiver should keep sampling signals in the channel with the frequency twice as the symbol rate of the transmission. Let $s[1]$, $s[2]$, $s[3] \ldots$ denote samples obtained in sampling procedure. To detect the beginning of a packet, the receiver calculate the correlation between pilot sequence and these samples as
\begin{equation}
S[i]=\sum_{k=0}^{L-1}~s[i+2k]p[k+1]
\end{equation}
When $S[i]$ and $S[i+1]$ spike consecutively, we can conclude that a packet arrives and $s[i]$ is the first sample for this packet. Similarly, we can locate the ending of a packet in the same way. Also, we note that when the preamble of a packet is corrupted by other packets, the method still works to indicate the beginning of the packet. This is due to the fact that the correlation between pilot sequence and an irrelevant sequence is expected much less than that between the pilot sequence and itself, and hence the existence of components of other packet in samples has less impact on the occurrence of the correlation peak indicating the arrival of the packet. Therefore the receiver is able to detect the beginning of each packet from superposed waveform with correlation based method.

\paragraph{Channel Estimation}
Once detecting the arrival of the superposed waveform broadcast from the relay node, the receiver of the end node needs to estimate the channel gains experienced by both packets. If the channel estimations are performed respectively for two packets considering another packet as interference, the accuracy of the estimations are poor due to the existence of strong interferer. Therefore we jointly estimate the channel coefficients for two packets, which is similar with idea in \cite{CEM}.

For the sake of clarity, we name two packets as packet F and packet S according to their transmission order. Let $h_{F}$ and $h_{S}$ denote the fading coefficients experienced by packet F and packet S, $c_{F}[n] \ (1 \leqslant n \leqslant L)$ and $c_{S}[n] \ (1 \leqslant n \leqslant L)$ represent the symbol sequences for two packets, $g_{F}(t)$ and $g_{S}(t)$ stand for the distorted pulse shapes after transmission, and $T_{d}$ denote the relative delay between two packets. Also, $w(t)$ is the noise process. We should emphasize that $h_{F}$, $h_{S}$, $g_{F}(t)$, $g_{S}(t)$ and $w(t)$ all count not only the contribution due to the up-link from the end node to the relay node but also that due to the downlink from the relay node to the end node. Then the received waveform $y(t)$ at the end node can be expressed
\begin{equation}
y(t)=\sum_{n=1}^{L}h_{F}c_{F}[n]g_{F}(t-(n-1)T)+\sum_{n=1}^{L}h_{S}c_{S}[n]g_{S}(t-(n-1)T-T_{d})+w(t)
\end{equation}
After sampling process, the $i$th sample is given by
\begin{equation}
y[i]=\sum_{n=1}^{L}h_{F}c_{F}[n]g_{F}((i-1)\frac{T}{2}+\Delta-(n-1)T)+\sum_{n=1}^{L}h_{S}c_{S}[n]g_{S}((i-1)\frac{T}{2}+\Delta-(n-1)T-T_{d})+ w((i-1)\frac{T}{2}+\Delta)
\end{equation}
Where $\Delta$ is the time at the first sampling position of packet F and we have that $0 \leqslant \Delta \leqslant \frac{T}{2}$. Considering odd-index samples, we have
\begin{eqnarray} \label{oddsample}
y_{odd}[k] & = & y[2k-1] \\
           & = & \sum_{n=1}^{L}h_{F}c_{F}[n]g_{F}(kT+\Delta-nT)+\sum_{n=1}^{L}h_{S}c_{S}[n]g_{S}(kT+\Delta-nT-T_{d})+w(kT-T+\Delta) \\
           & = & \sum_{n=1}^{L}h_{F}c_{F}[n]g_{F}(kT+\Delta-nT)+\sum_{n=1}^{L}h_{S}c_{S}[n]g_{S}(kT+\delta-nT-DT)+w(kT-T+\Delta)
\end{eqnarray}
Where $D=\lfloor (T_{d}-\Delta)/T \rfloor$ and $\delta=T_{d}-\Delta-DT$ (without loss of generality, we can assume that $0 \leqslant \delta \leqslant \frac{T}{2}$). In addition, let $[0,L_{h}T]$ denote the region\footnote{The length of this region is close to the delay spread $\tau$ of the channel, which is about a few symbol times in real systems as indicated by \cite{delayspread}. } where the value of $g_{v}(t) \ (v\in\{S,F\})$ is evidently above the level of noise strength. To simplify the equation (\ref{oddsample}), we can assume that
\begin{equation}
g_{v}(t)=0
\end{equation}
when $t<0$ or $t>L_{h}T$. Hence, equation (\ref{oddsample}) can be simplified as
\begin{eqnarray*}
y_{odd}[k] & = & \sum_{n=k-L_{h}+1}^{k}h_{F}c_{F}[n]g_{F}(\Delta+(k-n)T) +
                 \sum_{n=k-D-L_{h}+1}^{k-D}h_{S}c_{S}[n]g_{S}(\delta+(k-D-n)T)\\
           & &   +    w(kT-T+\Delta) \\
           & = & \sum_{n=0}^{L_{h}-1}h_{F}c_{F}[k-n]g_{F}(\Delta+nT)+ \sum_{n=0}^{L_{h}-1}h_{S}c_{S}[k-n-D]g_{S}(\delta+nT)++w(kT-T+\Delta)
\end{eqnarray*}
The previous equation can be written in matrix form as
\begin{equation}
y_{odd}=[C_{F} C_{S}]
\left[
\begin{array}{c}
h_{F,odd} \\
h_{S,odd}
\end{array}
\right]
+w_{odd}
\end{equation}
where $h_{F,odd}$ and $h_{S,odd}$ are matrixes with dimension $L_{h}\times 1$ and
\begin{equation}
h_{F,odd}= h_{F}\mathbf{g}_{F,odd}=h_{F}
\left[
\begin{array}{c}
g_{F}(\Delta) \\
g_{F}(\Delta+T) \\
\vdots \\
g_{F}(\Delta+L_{h}T) \\
\end{array}
\right]
\end{equation}
\begin{equation}
h_{S,odd}= h_{S}\mathbf{g}_{S,odd}=h_{S}
\left[
\begin{array}{c}
g_{S}(\delta) \\
g_{S}(\delta+T) \\
\vdots \\
g_{S}(\delta+L_{h}T) \\
\end{array}
\right]
\end{equation}
$C_{F}$ and $C_{S}$ are matrixes with the dimension $(L+D) \times L_{h}$ and
\begin{equation}
C_{F}=\left[
\begin{array}{cccc}
c_{F}[1] & 0 & \cdots & 0 \\
c_{F}[2] & c_{F}[1] & \cdots & 0 \\
\vdots & \vdots &  \ddots & \vdots \\
c_{F}[L] & c_{F}[L-1] & \cdots & c_{F}[L-L_{h}+1] \\
0 & c_{F}[L] & \cdots &  c_{F}[L-L_{h}+2] \\
\vdots & \vdots & \ddots & \vdots \\
0 & 0 & \cdots & c_{F}[L] \\
0 & 0 & \cdots & 0 \\
\vdots & \vdots & \ddots & \vdots \\
0 & 0 & \cdots & 0
\end{array}
\right]
\end{equation}
\begin{equation}
C_{S}=\left[
\begin{array}{cccc}
0 & 0 & \cdots & 0 \\
\vdots & \vdots & \ddots & \vdots \\
0 & 0 & \cdots & 0 \\
c_{S}[1] & 0 & \cdots & 0 \\
c_{S}[2] & c_{S}[1] & \cdots & 0 \\
\vdots & \vdots &  \ddots & \vdots \\
c_{S}[L] & c_{S}[L-1] & \cdots & c_{S}[L-L_{h}+1] \\
0 & c_{S}[L] & \cdots &  c_{S}[L-L_{h}+2] \\
\vdots & \vdots & \ddots & \vdots \\
0 & 0 & \cdots & c_{S}[L] \\
\end{array}
\right]
\end{equation}
$w_{odd}$ is a matrix with the dimension $(L+D) \times 1$ and
\begin{equation}
w_{odd}=
\left[
\begin{array}{c}
w(\Delta) \\
w(T+\Delta) \\
\vdots \\
w((L+D-1)T+\Delta)
\end{array}
\right]
\end{equation}
Similarly, even-index samples $y_{even}$ can be denoted as
\begin{equation}
y_{even}=[C_{F} C_{S}]
\left[
\begin{array}{c}
h_{F,even} \\
h_{S,even}
\end{array}
\right]
+w_{even}
\end{equation}
where $h_{F,even}$, $h_{S,even}$ and $w_{even}$ is given by
\begin{equation}
h_{F,even}=h_{F}\mathbf{g}_{F,even}=h_{F}
\left[
\begin{array}{c}
g_{F}(\Delta+T/2) \\
g_{F}(\Delta+T/2+T) \\
\vdots \\
g_{F}(\Delta+T/2+L_{h}T) \\
\end{array}
\right]
\end{equation}
\begin{equation}
h_{S,even}=h_{S}\mathbf{g}_{S,even}=h_{S}
\left[
\begin{array}{c}
g_{S}(\delta+T/2) \\
g_{S}(\delta+T/2+T) \\
\vdots \\
g_{S}(\delta+T/2+L_{h}T) \\
\end{array}
\right]
\end{equation}
and
\begin{equation}
w_{oven}=
\left[
\begin{array}{c}
w(T/2+\Delta) \\
w(T+T/2+\Delta) \\
\vdots \\
w((L+D-1/2)T+\Delta)
\end{array}
\right]
\end{equation}
Consider the first 2$L_{p}$ samples $y_{1}$ and the last 2$L_{p}$ samples $y_{2}$, i.e. samples which coincide with the preamble of packet F and the postamble of packet S, as shown in Figure \ref{jdec}. Then odd-index samples in $y_{1}$ and $y_{2}$ can be expressed as
\begin{eqnarray} \label{esteqn}
\left[
\begin{array}{c}
y_{1,odd} \\
y_{2,odd}
\end{array}
\right]
& = &
\left[
\begin{array}{cc}
C_{F,1} & C_{S,1} \\
C_{F,2} & C_{S,2}
\end{array}
\right]
\left[
\begin{array}{c}
h_{F,odd} \\
h_{S,odd}
\end{array}
\right]+
\left[
\begin{array}{c}
w_{1,odd} \\
w_{2,odd}
\end{array}
\right] \\
 & = &
C_{est}
\left[
\begin{array}{c}
h_{F,odd} \\
h_{S,odd}
\end{array}
\right]+
\left[
\begin{array}{c}
w_{1,odd} \\
w_{2,odd}
\end{array}
\right]
\end{eqnarray}
where $C_{F,1}$ consists of the first $L_{p}$ rows of $C_{F}$ and $C_{F,2}$ consists of the last $L_{p}$ rows of $C_{F}$. $C_{S,1}$, $C_{S,2}$, $w_{1,odd}$ and $w_{2,odd}$ are defined in a similar way. We note that entries in the first $L_{p}$ rows and the last $L_{p}$ rows of the matrix $C_{F}$ and $C_{S}$ only involves symbols in the preamble or the postamble and hence are known by the receiver. Therefore the matrix $C_{est}$ is
fully known and we can use Equation (\ref{esteqn}) to estimate the channel coefficients $h_{F,odd}$ and $h_{S,odd}$. Based on the Least Square Estimation, we have
\begin{equation}
\left[
\begin{array}{c}
\tilde{h}_{F,odd} \\
\tilde{h}_{S,odd}
\end{array}
\right] =
(C_{est}^{H}C_{est})^{-1}C_{est}^{H}
\left[
\begin{array}{c}
y_{1,odd} \\
y_{2,odd}
\end{array}
\right]
\end{equation}
Similarly, it can be shown that
\begin{equation}
\left[
\begin{array}{c}
\tilde{h}_{F,even} \\
\tilde{h}_{S,even}
\end{array}
\right] =
(C_{est}^{H}C_{est})^{-1}C_{est}^{H}
\left[
\begin{array}{c}
y_{1,even} \\
y_{2,even}
\end{array}
\right]
\end{equation}
We note that the inverse matrix of $C_{est}^{H}C_{est}$ exists if and only if the matrix $C_{est}$ has full rank. If the same set of pilot sequences, including the preamble and the postamble, is chosen by packet F and packet S, it is required that two packets have relative delay at least $L_{h}$ symbol times to guarantee the full rank of  $C_{est}$. As mentioned previously, this time period is close to delay spread of the channel and is on the level of nanoseconds. Hence the relative delay between two packets within this time duration can be easily avoided through deliberately introducing delay by the MAC layer when necessary. Another solution is to use different sets of pilot sequences for packet F and packet S. In this case, it can be shown that the rank of $C_{est}$ is always full, and hence the MAC layer only need to establish the rules about how to choose different sets of pilot sequences in different situations. This solution incurs less complexity on the MAC layer design and also avoid the overhead due to the deliberately introduced delay in previous solution. However, this scheme doubles the computational load for packet detecting, i.e. the receiver has to correlate signals samples with four sequences (preamble sequences and postamble sequences for packet F and packet S individually) to detect the beginning and the ending of packets. The better solution is to choose the same set of pilot sequences for packet S and packet F but to use sequences with different order, i.e. the preamble of packet F is the postamble of packet S while the postamble of packet F is the preamble of packet S. By this mean, the receiver only needs to correlate signal samples with two sequences\footnote{Known that two packets overlap with each other, this operations would not result in the confusion between the ending of packet F and the beginning of packet S.} and hence the computational load for packet detecting keeps unchanged. At the same time, similar with the previous scheme this solution can guarantee the full rank of the matrix $C_{est}$ in any cases. Hence due to its several advantages, this solution is adopted in most scenarios.

In the end of this section, we emphasize that channel coefficients $h_{v,u} \ (v \in \{F,S\}, ~ u \in\{odd,even\})$ count not only fading terms $h_{F}$ (or $h_{S}$) but also the pulse-shape gain $g_{F}(\Delta+nT)$ (or $g_{S}(\delta+nT)$). In general decoding procedure, with timing synchronization the sampling offset $\Delta$ is fixed to a certain value corresponding to the optimal sampling point, and hence the pulse shape gain is constant. However, when two packets superpose together but without any coordination, their relative delay $T_{d}$ may change from transmission to transmission, and hence so do $\Delta$ and $\delta$. Therefore we cannot assume the channel coefficients are constant even in slow fading channel and have to estimate it every time.

\paragraph{Self-packet Identification} The receiver also need to determine which packet, packet F or packet S, is transmitted by itself. This can be achieved by correlating waveform samples with the symbol sequence $c_{k}[n]$ of the known packet. Because that 2X oversampling is adopted, every symbol would be sampled twice. Hence $y[2n-1]$ and $y[2n]$ correspond to the same symbol in packet F. Therefore if the symbol sequence of known packet align with the position of packet F in superposed waveform the correlation can be calculated as
\begin{equation}
Corr_{F,1}=\sum_{n=1}^{L} y[2n-1]c_{k}[n]
\end{equation}
and
\begin{equation}
Corr_{F,2}=\sum_{n=1}^{L} y[2n]c_{k}[n]
\end{equation}
Then we can use $R_{F}=max\{Corr_{F,1},Corr_{F,2}\}$ to indicate the relevance between the packet F and known packet. Similarly, if $i_{0}$ denote the first sample of packet S, when the symbol sequence of known packet align with the position of packet S in superposed waveform the correlation can be calculated as
\begin{equation}
Corr_{S,1}=\sum_{n=1}^{L} y[(i_{0}-1)+2n-1]c_{k}[n]
\end{equation}
and
\begin{equation}
Corr_{S,2}=\sum_{n=1}^{L} y[(i_{0}-1)+2n]c_{k}[n]
\end{equation}
Also
\begin{equation}
R_{S}=max\{Corr_{S,1},Corr_{S,2}\}
\end{equation}
If $R_{F} \geqslant R_{S}$, the relevance between packet F and known packet is higher than that between packet S and known packet. Hence the receiver decide that packet F is known packet while packet S is desired packet. However, if $R_{F} < R_{S}$, the decision is reversed. Without loss of generality, we can assume that packet F is already known by the receiver and packet S needs to be decoded.

\paragraph{Waveform Recovery} Based on previous knowledge, we remove the components due to known packet from waveform samples to enable the decoding for desired packet. The processed samples are denoted as $y_{S,odd}$ (odd-index samples) and $y_{S,even}$ (even-index samples), and they are given by
\begin{eqnarray}\label{ysodd}
y_{S,odd} & = & y_{odd}-C_{F}\tilde{h}_{F,odd} \\
          & = & C_{S}h_{S,odd}+C_{F}(h_{F,odd}-\tilde{h}_{F,odd})+w_{odd} \\
          & = & C_{S}h_{S,odd}+ \tilde{w}_{odd}
\end{eqnarray}
and
\begin{eqnarray}\label{yseven}
y_{S,even} & = & y_{even}-C_{F}\tilde{h}_{F,even} \\
          & = & C_{S}h_{S,even}+C_{F}(h_{F,even}-\tilde{h}_{F,even})+w_{even} \\
          & = & C_{S}h_{S,odd}+ \tilde{w}_{even}
\end{eqnarray}
Up to now, we can decode packet S based on Equation (\ref{ysodd}) or Equation (\ref{yseven}). However, due to the uncertainty of $\Delta$ and $\delta$, the pulse shape gain $\mathbf{g}_{F}$ (or $\mathbf{g}_{S}$) may not be optimal. This results in the degradation of signal-noise ration, which is proportional to $|h_{S,u}|^{2}/E[w_{u}^{2}]=|h_{S}^{2}||\mathbf{g}_{S,u}|^2/E[w_{u}^{2}] \ (u \in \{odd,even\})$, and hence has negative impact on decoding performance. The better solution is to recovery the waveform of packet S taking advantage of redundant samples and relocate the optimal sampling points. For convenience, we emerge $y_{S,odd}$ and $y_{S,even}$ together as
\begin{equation}
y_{S}[k]=\left\{
\begin{array}{ll}
y_{S,odd}[(k+1)/2] & \textrm{k is odd} \\
y_{S,even}[k/2] & \textrm{k is even}
\end{array}
\right.
\end{equation}
Then the waveform of packet S can be recovered as
\begin{equation}
\tilde{y}_{S}(t)=\sum_{n=1}^{2L}y_{S}[n]sinc(\frac{t-n\frac{T}{2}}{\frac{T}{2}})
\end{equation}
In reality, the calculation of $\tilde{y}_{S}(t_{0})$ is approximated by the summation over a few items the indices of which are close to $2t_{0}/T$.
\paragraph{Decoding} The recovered waveform can be decoded with general decoding procedure. Hence we can consider all of previous steps as a preprocess unit before general decoding block which extracts the waveform of desired packet from superposed waveform. Also, this preprocess unit is independent of the decoding block following it. This feature make our algorithm applicable to a wide variety of modulation schemes.

\section{Two-way Relaying Enhanced Ad-hoc Network Protocol}
In this section, we propose a new protocol called TREAN (Two-way Relay Enhanced Ad-hoc Network) that incorporates our new physical-layer scheme into general ad-hoc network to boost the performance. The protocol is a random access MAC scheme in nature and borrows some essentials from CSMA/CA protocol. It performs RTS/CTS-like queries to build the cooperative configuration for two-way relaying. After that the new physical layer scheme is applied to conduct two-way relay transmission with least coordination. In the rest of this section, we first present the basic mode of TREAN protocol to provide a detailed description about how TREAN protocol works. Then two extended modes are discussed to further enhance the performance of TREAN protocol.


%
%
%
%
%
\subsection{The Basic Mode} \label{bmode}
\begin{figure}[ht]
\centering
\includegraphics[width=0.8\textwidth]{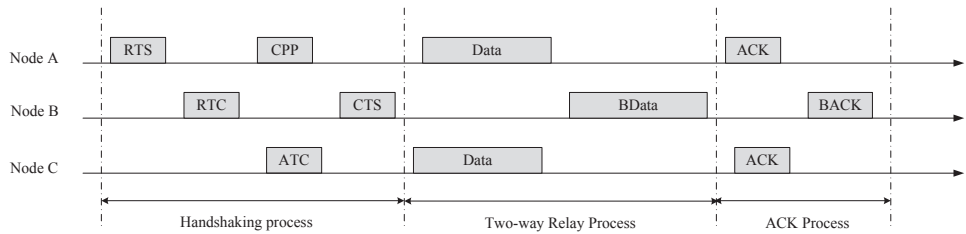}
\caption{\label{flow} The two-way relaying cooperation process in TREAN protocol}
\end{figure}

The two-way relaying cooperation process in TREAN protocol can be divided into three sub-processes, as shown in Figure \ref{flow}.
The handshaking process sets up the connections between stations for the cooperation and clears the channel for data transmissions. Following that the two-way relay process adopts our new physical layer technique to transmit data packets. Finally, the ACK process reports successful transmissions also with the help of our two-way relaying scheme. The detailed procedure are discussed as follow.

\begin{figure}[ht]
\centering
\includegraphics[width=0.2\textwidth]{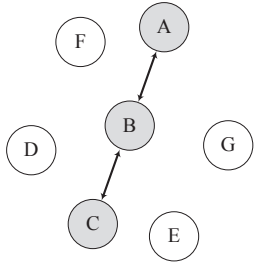}
\caption{\label{twr} The schematic diagram for illustrating the basic model of TREAN. The station $A$ has a packet with stations $B$ and $C$ as next two hops on its routing path and initiate the transmission by sending the RTS frame. Station $C$ is called next two-hop destination of the packet. The next two-hop destination is not necessarily the final destination of the packet.}
\end{figure}

\begin{figure}[ht]
\centering
\includegraphics[width=0.45\textwidth]{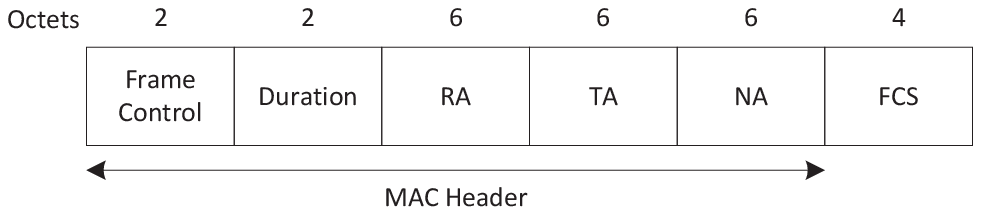}
\caption{\label{RTSframe} The format of the RTS frame in TREAN. The Frame Control field and FCS field are same as those specified in 802.11 Standard \cite{80211}. RA and TA are addresses of the receiver and the transmitter of the RTS frame. NA is the address of next two-hop destination of the data packet being transmitted.}
\end{figure}

\paragraph{Handshaking process} Consider a station $A$ with a data packet to transmit, as shown in Figure \ref{twr}.
As CSMA/CA protocol, the station sends the RTS frame first when the channel is sensed idle and the backoff time counter decreases to zero. The contents of the RTS frame includes those in its CSMA/CA counterpart. Besides, the address of next two-hop destination on the routing path of the data packet should be added in the RTS frame for TREAN protocol. The format of modified RTS frame is shown in Figure \ref{RTSframe}. If the source routing is used, we can extract the address of the next two-hop destination from data packets directly. If the hop-by-hop routing is used, to provide this information, the routing table at every station should indicate next two hops towards each possible destination. This can be achieved by exchanging the routing information between neighbor stations periodically. A special case is that the data packet would reach its final destination after next hop, i.e. the next two-hop destination does not exist for the packet. In this scenario, a special RTS are sent and the transmission procedure is same with CSMA/CA protocol.

Also, we modify the blocking function of the RTS frame\footnote{The blocking function of the RTC frame and the ATC frame is modified in similar way}. In TREAN protocol, the RTS frame only blocks neighbor stations in the period from the end of the transmission of itself to the time when the data frame should be transmitted. If the handshaking process is free from the interference and hence the data frame is transmitted, neighbor stations would keep inactive until the estimated ending time of the cooperation according to the NAV information carried by the data frame.
The reason for this modification is based on two facts. On the one hand, because the whole transmission process under TREAN protocol contains a two-way relaying cooperation, as shown in Figure \ref{flow}, the NAV time duration (without modification) in the RTS frame of TREAN protocol would be about two times larger as that in CSMA/CA protocol. On the other hand, the two-way relaying cooperation in TREAN protocol involves three stations, and this relatively complicated configuration is more vulnerable to hidden node problem which may result in the failure of the transmission. Therefore it is possible that the cooperation initiated by a RTS frame fails but all of neighbor of the transmitter of RTS are blocked for a long period. This has negative impact on the performance. With the modification, the blocking on neighbor stations would be removed after a short time period if the cooperation initiated by the RTS frame fails.

In addition, after the transmission of the RTS frame, a RTS-frame timer is set up. If no valid CTS frame are received before the timeout, the station increases its backoff stage and take a random backoff.

If the RTS frame are successfully sent to its receiver, the receiver would transmit a RTC (Request to Cooperate) frame to the station whose address is written in NA field of received RTS frame after waiting for SIFS period. For example, as shown in Figure \ref{twr}, if station $B$ are correctly received the RTS frame from station $A$, it would send RTC frame to station $C$. Also, this frame can be overheard by station $A$.
\begin{figure}[ht]
\centering
\includegraphics[width=0.65\textwidth]{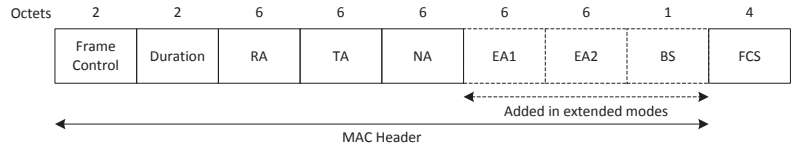}
\caption{\label{RTCframe} The format of the RTC frame in TREAN protocol. The Frame Control field and FCS field are same as those specified in 802.11 Standard \cite{80211}. RA and TA are addresses of the receiver and the transmitter of the RTC frame. NA is the address of the transmitter of RTS frame.}
\end{figure}

The format of the RTC frame is shown in Figure \ref{RTCframe}. The information carried by this frame includes not only the addresses of the transmitter and the receiver of this frame, but that of the sender of the previously received RTS frame. Also, the operation mode of TREAN protocol is specified by this frame. This is achieved by using different values in the subtype sub-field contained in frame control field of the frame, as shown in Table \ref{subtype}.
\begin{table}[ht]
\centering
\includegraphics[width=0.8\textwidth]{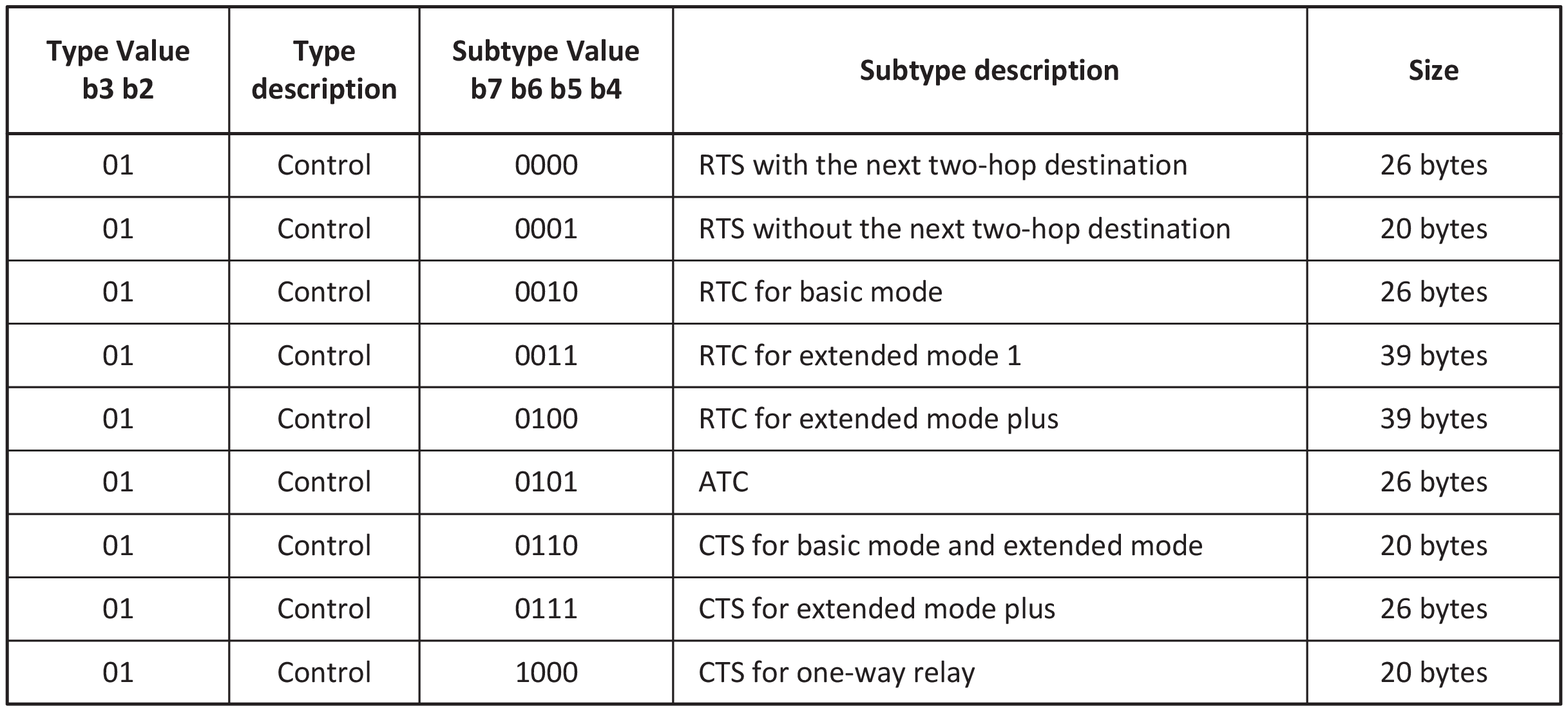}
\caption{\label{subtype}The subtype field for control frames in TREAN protocol. All of these combinations of type field and subtype field are reserved and not used in 802.11 standard \cite{80211}.}
\end{table}

\begin{figure}[ht]
\centering
\includegraphics[width=0.45\textwidth]{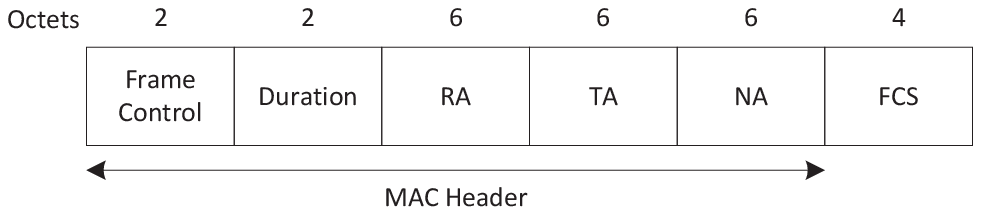}
\caption{\label{ATCframe} The format of the ATC frame in TREAN. The Frame Control field and FCS field are same as those specified in 802.11 Standard \cite{80211}. RA and TA are addresses of the receiver and the transmitter of the ATC frame. NA is the address copied from the NA field of the RTC frame.}
\end{figure}

If the RTC frame are correctly received by its receiver the station $C$, the address written in NA field of the RTC frame are extracted firstly. Then the station $C$ would check the transmission buffer to determine whether it has a data packet towards the station indicated by this address (the station $A$ in our example). If there does exist such a packet, the station $C$ would send back a ATC (Answer to Cooperate) frame to the transmitter of the RTC frame, namely station $B$. The format of the ATC frame is shown in Figure \ref{ATCframe}.

\begin{figure}[ht]
\centering
\includegraphics[width=0.6\textwidth]{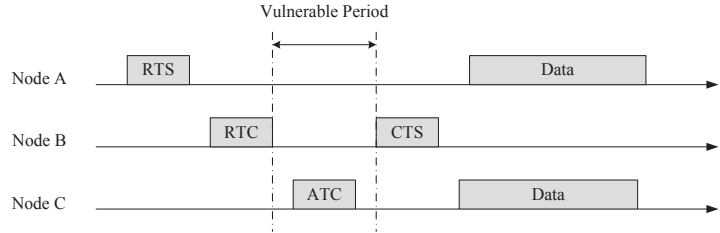}
\caption{\label{cpp} The vulnerable period for station $A$.}
\end{figure}

Also, the RTC frame can be overheard by station $A$. If station $A$ takes no action after receiving this frame, both station $A$ and station $B$ have no transmission until the broadcast of the CTS frame as shown in Figure \ref{cpp}. This may cause that the channel around the station $A$ is recaptured by other nodes due to no transmission in nearby region. The case is worse when the ATC frame is transmitted with a relatively low rate and hence the vulnerable period becomes longer. Also, if the station density is high, the recapture of the channel tends to happen more frequently.

The blocking function of the RTS frame in CSMA/CA protocol can only solve the issue partially. The stations outside the communication range but in the interference range of station $A$ are not affected by RTS frame, and they can recapture the channel and cause the failure of the cooperation initiated by station $A$.

The better solution for this problem is using CPP (Channel Protection Packet), i.e. the station $A$ sends a packet after receiving RTC frame to protect its channel, as shown in Figure \ref{flow}. Although solving previous issue, this may cause the signal superposition of ATC frame and CPP frame at station $B$. However, if the CPP frame is known by the station $B$, then our physical layer technique can be applied to cancel the CPP frame and extract the ATC frame.

We can choose the RTS frame previously transmitted by station $A$ as a CPP frame. Also, we require that the station $C$ changes the common order of pilot sequences for its ATC frame, namely swap the preamble and the postamble, to guarantee the function of channel estiamtion in our physical layer scheme. If all of these are done, station $B$ can retrieve buffered RTS frame and use it extract ATC frame with our physical layer decoding algorithm once the collision happens.

\begin{figure}[ht]
\centering
\includegraphics[width=0.45\textwidth]{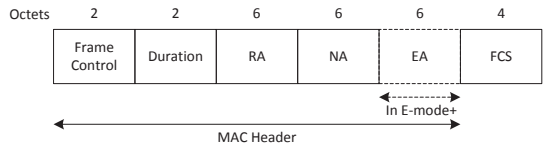}
\caption{\label{CTSframe} The format of the CTS frame in TREAN. The Frame Control field and FCS field are same as those specified in 802.11 Standard \cite{80211}. RA and TA are addresses of the transmitters of the RTS frame and the ATC frame respectively. Also, an EA field is added in extended mode plus to indicate the destination of backward data frame.}
\end{figure}

If the station $B$ receives the ATC frame before the timeout of RTCframe-timer, then it broadcasts a CTS frame to station $A$ and station $C$. This frame serves as an authorization for the two-way relay cooperation between station $A$ and station $C$ and at the same time blocks the transmissions of all other neighbors of station $B$. However, if no ATC frame received before the timeout, station $B$ presume that station $C$ does not has a data packet towards station $A$. In this case, station $B$ would broadcast a special CTS indicating that one-way relay are performed in following stages. The format of the CTS frame is shown in Figure \ref{CTSframe}.

\paragraph{Two-way relay process} If the CTS frame for two-way relay is received by station $A$ and station $C$, they would transmit their data packets to station $B$ after a SIFS time period. As the case of the ATC frame, station $C$ swaps the preamble and the postamble for its data frame. Then station $B$ amplifies and forwards the received waveform to the transmitters of two data packets. Once receiving the waveform, station $A$ and station $C$ can decode their desired data packets with our physical layer decoding algorithm. However, if the CTS frame for one-way relay is broadcast, then only station $A$ transmit its data packets to station $B$. Once receiving the packets, the station $B$ forwards the packets to the station $C$ after waiting for SIFS.

\paragraph{ACK process} If station $A$ and station $C$ decodes data packets correctly, they send ACK frames to announce the successful reception, and the transmission process is also in two-way relaying manner as shown in Figure \ref{flow}.

\begin{figure}[ht]
\centering
\includegraphics[width=0.57\textwidth]{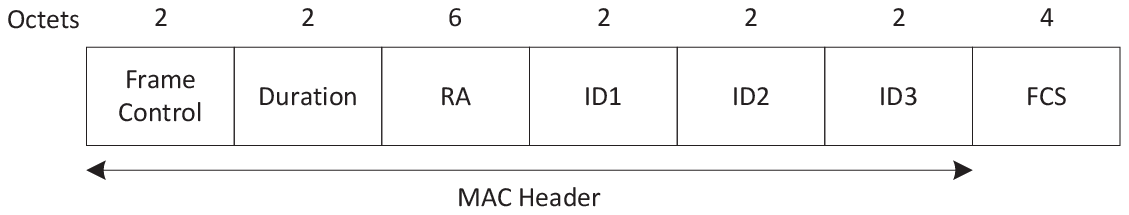}
\caption{\label{ACKframe} The format of the ACK frame in TREAN. The Frame Control field Duration field, FCS field are same as those specified in 802.11 Standard \cite{80211}. RA is addresses of the transmitter of the data. ID1, ID2 and ID3 are frame ID for most recently received data frames from RA.}
\end{figure}

Because the structure unit of our physical layer involves three nodes to cooperate, it is more complicated and larger than that of traditional physical layer. Hence our scheme suffers more serious hidden node problem than those traditional ones. One of effects of this issue is increasing the loss rate of ACK frames. Our solution to this problem is to acknowledge preivously received data frames from same sender in current ACK frame. This can be achieved by including frame IDs of recently received data frames in the ACK frame to notice the successful transmission of these frames. The format of the ACK frame in TREAN protocol is shown in Figure \ref{ACKframe}.
Under this ACK scheme, a station buffers the transmitted data frame instead of retransmitting immediately if the corresponding ACK frame is not received. Until the next ACK frame from the same transmitter of the lost ACK arrives, the station can determine whether previous data frame is received according to the ID fields in the ACK frame and then retransmit or delete the frame from the buffer.
\subsection{Extended Modes}
\paragraph{Extended Mode} The basic mode of TREAN protocol leads to an enormous improvement on the throughput of the network when the answer-to-cooperation(ATC) probability is high. The ATC probability is here defined as the probability that the station has data frames to send back towards the address in NA field of the RTC frame. However the performance degrades when the ATC probability goes down. Unfortunately, in many realistic systems, the ATC probability is low. To further improve the performance of TREAN protocol even in low ATC-probability scenario, extended mode of TREAN protocol is needed.

To run the extended mode of TREAN protocol, every station in the network should maintain a special neighbor table, called close neighbor table. The table contains neighbors which are close to the station in distance and hence have high-quality communication links with the station. The threshold for recognizing a close neighbor depends on the noise-level, the station density of the network, etc. In addition, stations should exchange the information of close neighbor tables with their all neighbors periodically.

\begin{figure}[ht]
\centering
\includegraphics[width=0.2\textwidth]{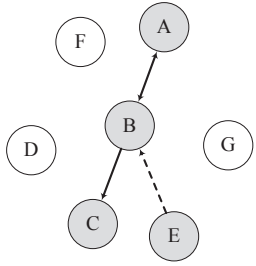}
\caption{\label{twr1} The schematic diagram for illustrating the extended mode of TREAN protocol. The station $A$ has a packet with stations $B$ and $C$ as next two hops on its routing path and initiate the transmission by sending the RTS frame. Station $E$ is a close neighbor of station $C$, and both of them may have data frames to send back to the station $A$.}
\end{figure}

The basic idea behind the extended mode of TREAN protocol is to provide the cooperation opportunity to more stations. As shown in Figure \ref{twr1}, we assume that station $C$ does not have a data frame towards the station $A$ but station $E$ has ones coincidentally. In the basic mode, the two-way relay cooperation cannot come into being due to no backward data frames in station $C$. However, if the cooperation opportunity can be offered to station $E$, then a special two-way relay cooperation can form to transmit the frame of station $A$ to station $C$ and at the same time send that of station $E$ to the station $A$ with the help of station $B$.

The offer of the cooperation opportunity to more stations are done by station $B$ with the help of the extended RTC frame, as shown in Figure \ref{RTCframe}. After receiving the RTS frame from station $A$, the station $B$ randomly chooses some stations from the close neighbor table of station $C$ (indicated by NA field of the RTS frame), and writes addresses of these stations in the EA fields of the extended RTC frame indicating that these stations are also allowed to participate in the cooperation beside the station $C$. For convenience, we call all of these stations (including stations $C$) as qualified stations. Evidently, the maximum possible number of qualified stations is equal to the number of EA fields $n_{EA}$ in an extended RTC frame plus one. Hence the station $B$ can randomly allocate a sequence number in the range from 0 to $n_{EA}$ to every qualified station and guarantee that no stations share the same sequence number. These allocated sequence numbers are written in the TS field of the extended RTC frame in the specified order and reflect the priority of qualified stations participating in the cooperation. Specifically, a qualified station with sequence number $i$ has to wait for the period equal to  $(\textrm{SIFS}+ i \times \textrm{TimeSlot})$
before its transmission of ATC frame. However, to avoid the interruption by irrelevant stations, the waiting time for qualified stations should not be longer than DIFS. This sets the limitation on the maximum sequence number $n_{EA}$, i.e. the maximum number of EA fields in a extended RTC frame should not be greater than $(\textrm{DIFS}-\textrm{SIFS})/\textrm{TimeSlot}$. If DIFS and SIFS are specified as those in 802.11 standard \cite{80211}, the number of EA fields should not be more than two. However, according to the simulation result, two EA fields is enough to enhance the throughput dramatically in low ATC-probability scenarios.

Once receiving the extended RTC frame, each qualified station would send its ATC frame after required waiting time if the station has data frames to send back to the station $A$. Before the transmission of ATC frames, every qualified station should keep sense the channel. If a strong signal is sensed, the qualified station can know that at least one of nearby stations is in the transmission state. However all of nearby stations except qualified ones has to wait at least DIFS after the transmission of the RTC frame. Therefore the qualified station can determine that another qualified station with less sequence number has transmitted a ATC frame to the station $B$, and hence cancel the ATC frame transmission itself. In this way, there exists only one ATC frame sent to the station $B$ unless that all qualified stations have no backward data frames towards station $A$. This avoids the collision of ATC frames from different qualified stations.

At the same time, the transmitter of the RTS frame, i.e. station $A$, overhears the transmission of the extended RTC frame. As the basic mode of TREAN protocol, once the RTC frame is received station $A$ transmits a CPP frame after a SIFS time period. Nevertheless, in the extended mode the transmission of the ATC frame could starts until $(\textrm{SIFS}+ i \times \textrm{TimeSlot})$ time period after the reception of the extended RTC frame as discussed previously. Hence the CPP frame and the ATC frame may superpose at the station $B$ with relative delay as large as a few slot times. In this case, OFDM-based or LCC-based physical layer techniques for two-way relaying cannot work because the asynchronization is beyond their limitation. Fortunately our physical layer scheme can deal with this case and help station $B$ extract the ATC frame from superposed waveform.

Once a ATC frame from qualified stations are received before the timeout of RTCframe-timer, the station $B$ broadcast a CTS frame containing the addresses of station $A$ and the transmitter of the received ATC frame. This indicates that the two stations can transmit their data frames in two-way relay process and others should keep from sending any information.

Without loss of generality, we can assume that station $E$ successfully send a ATC frame. Once receiving the CTS frame, station $E$ and station $A$ transmit their data frames to station $B$ after the time period equal to SIFS plus a tiny random delay. At the same time, the station $C$ should overhear the transmission of station $E$, although the frame is not targeted for it. However, due to the concurrent transmission of station $A$ with station $E$, the overhearing suffers the interference from station $A$. Nevertheless, this interference is negligible for two reasons. First, the direct link between station $A$ and station $C$ is weak. Otherwise it is not necessary to add a station $B$ in routing path between station $A$ and station $C$. Also, the signal strength from station $E$ to station $C$ is strong. This is due to the fact that station $E$, as a qualified station, is chosen from the close neighbor of station $C$. Therefore the signal from the station $A$ is easily surpassed by that from station $E$, and has less impact on the decoding of the overheared frame. If station $C$ obtain the data frame of station $E$, it can apply our physical technique to extract the data frame of station $A$ from superposed waveform received in the broadcast stage of two-way relay process. Also, station $A$ can extract its desired frame with our physical layer decoding algorithm. By this way, two data frames are delivered to their respective destinations. Then two stations relay ACK frames as the basic mode of TREAN protocol.

\paragraph{Extended Mode plus} If further improvement on the performance of TREAN protocol in low ATC-probability scenario is needed, we can increase the number of EA fields in an extended RTC frame to provide the cooperation opportunity to more stations and hence enhance the probability that two-way relay cooperation comes into being successfully. However, the number of EA fields is usually limited by the parameters SIFS, DIFS and time slot as mentioned previously. Another solution is to run the extended mode plus of TREAN protocol. The basic idea behind the extended mode plus is to relax the limitation on backward data frames, namely qualified stations can reply ATC frames not only when they have data frames to station $A$ but also when they have ones to a close neighbor of station $A$. This dramatically increases the probability of relaying ATC frames, and hence enhance the probability for successful formation of two-way relay cooperation. However this is at the cost of the management overhead, i.e. every station has to know close neighbor tables of its two-hop neighbors beside those of its direct neighbors.

\begin{figure}[ht]
\centering
\includegraphics[width=0.2\textwidth]{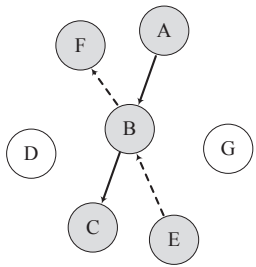}
\caption{\label{twr2} The schematic diagram for illustrating the extended mode plus of TREAN protocol. The station $A$ has a packet with stations $B$ and $C$ as next two hops on its routing path and initiate the transmission by sending the RTS frame. Station $E$ is a close neighbor of station $C$, and station $F$ is a close neighbor of station $A$}
\end{figure}

The runing of the extended mode plus can be further explained with the help of an example shown in Figure \ref{twr2}. The qualified staion $E$ has a data frame to station $F$, a close neighbor of station $A$. If no qualified stations with less sequence number have data frames to station $A$ or its close neighbor, station $E$ would transmit an ATC frame after waiting required time period. Once authorized by the CTS frame, station $E$ and station $A$ send their data frames to the station $B$. Meanwhile, the station $C$ and the station $F$ overhear transmissions from their respective close neighbors. After that the station $B$ would amplify and forward received superposed waveform to the station $C$ and the station $F$, and then they can use their overheard frame to extract their desired frames from the waveform with our physical layer scheme .

At the end of this section, we emphasize that although extended modes provide better performance in low-ATC-probability scenario, they would also result in larger management overhead. The extended mode requires stations to maintain close neighbor tables and exchange them with all neighbor stations. In extended mode plus, close neighbor tables of two-hop neighbors are further required to be known by stations. However, all of these are nor prerequisite in the Basic mode. Therefore we should select suitable mode according to different application scenario. In mobile network or the network with high bi-directional data flows, the basic mode is better solution, while in stationary network with low ATC-probability, extended modes are needed to enhance the performance. Also, the adaptive switch between different modes are feasible.
\section{Performance Analysis}
In this section, we adopt Markov model to derive the saturation throughput of the network with TREAN protocol. The objective is to evaluate the performance of TREAN protocol in theory. Also, the theoretical results can provide insights and guidelines for selecting optimal parameters in protocol design.

As prior works\cite{bianchi}, we define the throughput as the successfully transmitted payload in the network within unit time period. However, in this paper, we redefine the saturation condition as every station always has packets to transmit and each station has packets to send back when cooperative request is received, namely the transmission queue for any potential two-hop destinations are always non-empty. Also, to show the full capacity of TREAN protocol, we assume that all of transmissions are in two-way relaying manner, i.e. special RTS for the data frame without next two-hop destination is never initiated.

The derivation is divided into two subsections according to the size of networks where TREAN protocol is applied. In subsection \ref{ssd}, we derive the saturation throughput for small-scale network. Stations in this type of network can sense transmissions from all other stations, namely the network is free from the hidden node problem. In subsection \ref{mhop}, we provide an derivation of saturation throughput of the network deployed in large area. In this scenario, the size of the network is beyond the sensing range of the station, and hence the performance of the network would be affected by hidden stations.

\subsection{Saturation Throughput in small-scale network}\label{ssd}
Consider $n$ stations in a small-scale network. For every station, it may stay in different backoff stages and have different values in its backoff-time counter. Let $S_{i,j}$ denote the state that the station belongs to the $i$th backoff stage and have the value of the backoff counter equal to j. Note that the backoff stage is upper bound by constant $m$ and the value of backoff counter in stage $i$ should be in range from zero to contention window $W_{i}$ minus one.

There exists two important probabilities affecting transitions between different states $\{ S_{i,j} \}$. One of them is the transmission failure probability $p_{f}$ defined as the probability that collision happens in transmission process. As mentioned in \cite{bianchi}, it is reasonable to assume that $p_{f}$ is uncorrelated to the number of retransmission. Another key probability is the cooperation probability $p_{c}$, defined as the probability that a station accept to a two-way cooperation request and transmit a packet successfully in the cooperation. We assume that the cooperation probability is independent from backoff stages and values of backoff time counter due to the fact that the cooperation process is less related to the backoff behavior.

Based on previous discussion and inspired by prior work \cite{bianchi}, we can model the transitions between states $\{ S_{i,j} \}$ as a discrete-time Markov chain.In our Markov chain, all of non-zero state transition probabilities are:
\begin{equation}\label{tp}
\left\{
\renewcommand{\arraystretch}{1.8}
\begin{array}{ll}
p\{S_{0,j}|S_{i,0}\}=\displaystyle\frac{1-p_{f}}{W_{0}} & 0 \leqslant i \leqslant m, 0 \leqslant j \leqslant W_{0}-1 \\
p\{S_{i+1,j}|S_{i,0}\}=\displaystyle\frac{p_{f}}{W_{i+1}} & 0 \leqslant i \leqslant m-1, 0 \leqslant j \leqslant W_{i+1}-1 \\
p\{S_{m,j}|S_{m,0}\}=\displaystyle\frac{p_{f}}{W_{m}} & 0 \leqslant j \leqslant W_{m}-1 \\
p\{S_{0,k}|S_{i,j}\}=\displaystyle\frac{p_{c}}{W_{0}} & 0 \leqslant i \leqslant m, 0 \leqslant j \leqslant W_{i}-1, 0 \leqslant k \leqslant W_{0}-1  \\
\displaystyle p\{S_{i,j-1}|S_{i,j}\}=1-p_{c} & 0 \leqslant i \leqslant m, 1 \leqslant j \leqslant W_{i}-1 \\
\end{array}
\right.
\end{equation}
The first three equations in (\ref{tp}) correspond to backoff behavior after collision or successful transmission, which is similar with CSMA/CA protocol. However, the last two equations are special in the network adopting TREAN protocol. The fourth equation describes the state transition due to being requested to participate in the two-way relay cooperation. Once a station answers to cooperate in two-way relay manner and transmits a data frame successfully in the cooperation, the station reset its beckoff stage to zero and take a random backoff, just as a successful transmission under CSMA/CA protocol. The last equation in (\ref{tp}) accounts for the fact that as long as a station does not involve in a successful two-way relay cooperation it would reduce its backoff counter when channel is sensed idle for a certain period.

To determine the stationary distribution $\{ v_{i,j}\}$, we note that
\begin{equation}\label{std}
\left\{
\renewcommand{\arraystretch}{1.8}
\begin{array}{ll}
\displaystyle v_{0,j}=v_{0,j+1}(1-p_{c})+\frac{v_{c}}{W_{0}}+v_{t}\frac{1-p_{f}}{W_{0}} & 0 \leqslant j \leqslant W_{0}-1 \\
\displaystyle v_{i,j}=v_{i,j+1}(1-p_{c})+\frac{v_{i-1,0}p_{f}}{W_{i}} & 1 \leqslant i \leqslant m-1, \leqslant j \leqslant W_{i}-1 \\
\displaystyle v_{m,j}=v_{m,j+1}(1-p_{c})+\frac{v_{m-1,0}p_{f}}{W_{m}}+\frac{v_{m,0}p_{f}}{W_{m}} & 0 \leqslant j \leqslant W_{m}-1 \\
\sum_{i=0}^{m}\sum_{j=0}^{W_{i}-1}v_{i,j}=1 & \\
\end{array}
\right.
\end{equation}
where
\begin{equation}
\left\{
\renewcommand{\arraystretch}{1.8}
\begin{array}{l}
\displaystyle v_{c}=\left( \sum_{i=0}^{m}\sum_{j=1}^{W_{i}-1}v_{i,j}\right) \\
\displaystyle v_{t}=\sum_{i=0}^{m}v_{i,0}
\end{array}
\right.
\end{equation}
Based on recursive relations in equation (\ref{std}), we can show that
\begin{equation} \label{vij}
v_{i,j}=-\frac{a_{i}}{p_{c}}(1-p_{c})^{W_{i}-j}+\frac{a_{i}}{p_{c}}
\end{equation}
where
\begin{equation} \label{ai}
a_{i}=\left\{
\renewcommand{\arraystretch}{1.8}
\begin{array}{ll}
\displaystyle\frac{v_{c}}{W_{0}}+\displaystyle\frac{v_{t}(1-p_{f})}{W_{0}} & i=0 \\
\displaystyle\frac{v_{i-1,0}p_{f}}{W_{i}} & 1 \leqslant i \leqslant m-1 \\
\displaystyle\frac{v_{m-1,0}p_{f}}{W_{m}}+\frac{v_{m-1,0}p_{f}}{W_{m}} & i=m
\end{array}
\right.
\end{equation}
Substitute equation (\ref{ai}) into equation (\ref{vij}) and choose $j=0$, we can obtain new recursive equations about $\{ v_{i,0}\}$:
\begin{equation}
v_{i,0}=\left\{
\renewcommand{\arraystretch}{1.8}
\begin{array}{ll}
\displaystyle [1-(1-p_{c})^{W_{0}}][\frac{v_{c}}{W_{0}}+\frac{v_{t}(1-p_{f})}{W_{0}}]\frac{1}{p_{c}} & i=1 \\
\displaystyle \frac{p_{f}}{p_{c}W_{i}}[1-(1-p_{c})^{W_{i}}] \ v_{i-1,0} & 1 \leqslant i \leqslant m-1 \\
\displaystyle \frac{p_{f}[1-(1-p_{c})^{W_{m}}]}{p_{c}W_{m}-p_{f}[1-(1-p_{c})^{W_{m}}]} \ v_{m-1,0} & i=m-1
\end{array}
\right.
\end{equation}
Therefore, $v_{i,0}$ can be expressed as
\begin{equation}
v_{i,0}=\frac{p_{f}^{i}[v_{c}+v_{t}(1-p_{f})]}{p_{c}^{i+1}\delta_{i}}\prod_{k=0}^{i}\frac{1-(1-p_{c})^{W_{k}}}{W_{k}}
\end{equation}
Where
\begin{equation}
\delta_{i}=\left\{
\begin{array}{ll}
\displaystyle \frac{p_{c}W_{m}-p_{f}[1-(1-p_{c})^{W_{m}}]}{p_{c}W_{m}} & i=m \\
\displaystyle 1 & otherwise
\end{array}
\right.
\end{equation}
Hence, we have that
\begin{eqnarray}
v_{t} & = & \sum_{i=0}^{m}v_{i,0} \\
      & = & \sum_{i=0}^{m}\frac{p_{f}^{i}[v_{c}+v_{t}(1-p_{f})]}{p_{c}^{i+1}\delta_{i}}\prod_{k=0}^{i}\frac{1-(1-p_{c})^{W_{k}}}{W_{k}} \\
      & = & [v_{c}+v_{t}(1-p_{f})]\sum_{i=0}^{m}\frac{p_{f}^{i}}{p_{c}^{i+1}\delta_{i}}\prod_{k=0}^{i}\frac{1-(1-p_{c})^{W_{k}}}{W_{k}}
\end{eqnarray}
define
\begin{equation}
c(p_{f},p_{c})=\sum_{i=0}^{m}\frac{p_{f}^{i}}{p_{c}^{i+1}\delta_{i}}\prod_{k=0}^{i}\frac{1-(1-p_{c})^{W_{k}}}{W_{k}}
\end{equation}
Then we have that
\begin{equation}
v_{t}=[v_{c}+v_{t}(1-p_{f})]c(p_{f},p_{c})
\end{equation}
According to the definition of $v_{c}$ and $v_{f}$ and last equation in (\ref{std}), it can be shown that
\begin{equation}
v_{f}=[(1-v_{t})p_{c}+v_{t}(1-p_{f})]c(p_{f},p_{c})
\end{equation}
Moving $v_{t}$ to one side of the equality, we have that
\begin{equation}
v_{t}=\frac{c(p_{f},p_{c})p_{c}}{1-c(p_{f},p_{c})(1-p_{c}-p_{f})}
\end{equation}
Hence the transmission probability, defined as the probability that a station initiates the transmission in a randomly given time slot, can be expressed as
\begin{equation} \label{pt}
p_{t}=\sum_{i=0}^{m} \ v_{i,0}=v_{t}=\frac{c(p_{f},p_{c})p_{c}}{1-c(p_{f},p_{c})(1-p_{c}-p_{f})}
\end{equation}

For the sake of simplicity, we consider a symmetric setting where every station has equal opportunity to be requested to participate in the two-way relaying cooperation. In this case, we can assume that the cooperation probability $p_{c}$ is constant over stations in the network. Also, as mentioned in \cite{bianchi}, it is reasonable to assume that $p_{f}$ keeps invariant for all stations. Furthermore, we note that the transmission probability is decided by the cooperation probability $p_{c}$ and the transmission failure probability $p_{f}$. Hence we can conclude that the transmission probability $p_{t}$ is also unchanging for all stations.

Let $D_{X}$ denote the set of stations that have data frames with station X as their next two-hop destination and $T_{X}$ represent the set of stations which are next two-hop destinations of data frames from station X. Also, $p_{XY}^{t}$ stands for the probability that station $X$ transmits a RTS frame with the address of station Y in NA field, and $p_{XY}^{s}$ denotes the probability that such RTS frame is free from the collision and the cooperation process is successful. Then for a specific station $A$, the cooperation probability is given by
\begin{eqnarray}
p_{c} & = & \sum_{X \in D_{A}} \ p_{XA}^{t} p_{XA}^{s} \\
      & = & \sum_{X \in D_{A}} \ p_{XA}^{t} (1-p_{t})^{n-2} \\
      & = & (1-p_{t})^{n-2} \sum_{X \in D_{A}} \ p_{XA}^{t}
\end{eqnarray}
We add the cooperation probability for all stations together, then we have that
\begin{eqnarray}
\sum_{Y}p_{c} & = & \sum_{Y} (1-p_{t})^{n-2} \sum_{X \in D_{Y}} \ p_{XY}^{t} \\
              & = & (1-p_{t})^{n-2} \sum_{Y} \sum_{X \in D_{Y}} \ p_{XY}^{t} \\
              & = & (1-p_{t})^{n-2} \sum_{X} \sum_{Y \in T_{X}} \ p_{XY}^{t} \\
              & = & (1-p_{t})^{n-2} \sum_{X} p_{t}
\end{eqnarray}
We know that $p_{c}$ and $p_{t}$ keep constants for all stations, hence it can be shown that
\begin{equation}\label{pc}
p_{c}=p_{t}(1-p_{t})^{n-2}
\end{equation}
As derived in (\cite{bianchi}), the transmission probability can be expressed as
\begin{equation}\label{pf}
p_{f}=1-(1-p_{t})^{n-1}
\end{equation}
Combine equation (\ref{pt})  (\ref{pc}) (\ref{pf}), we can solve $p_{t}$,$p_{f}$ and $p_{c}$ with numerical methods. Although previous derivation is based on symmetric setting assumption, we should emphasize that our theoretical framework is not limited by this requirement. In fact, it can be applied to various scenarios. However, unsymmetrical scenarios results in relatively high computational complexity. For most general case, $p_{t}$,$p_{f}$ and $p_{c}$ are all different for different stations and we should solve $3n$ equations to obtain these probabilities for each station. In this case, some sophisticated numerical methods should be applied. For example, the secant updating methods can solve our problem at the cost of $\rm(n_{2})$ operations \cite{compute}.

Based on previous results, the length of generalized time slot $X$, defined as the time interval between two consecutive backoff counter decrements in \cite{bianchi}, can be calculated as
\begin{equation}
X=P_{idle}T_{slot}+P_{col}T_{c}+P_{succ}T_{s}
\end{equation}
Where $P_{idle}$, $P_{succ}$ and $P_{col}$ represents the probabilities that the channel is idle, captured by a successful transmission and occupied by a collision respectively. Also, $T_{slot}$, $T_{succ}$ and $T_{col}$ denotes the length of a time slot, a successful transmission period and time duration of a collision respectively. These variables are given by
\begin{equation}
\left\{
\renewcommand{\arraystretch}{1.8}
\begin{array}{l}
\displaystyle P_{idle}=(1-p_{t})^{n} \\
\displaystyle P_{succ}=n p_{t}(1-p_{t})^{n-1} \\
\displaystyle P_{col}=1-(1-p_{t})^{n}-n p_{t}(1-p_{t})^{n-1}
\end{array}
\right.
\end{equation}
and
\begin{equation}
\left\{
\renewcommand{\arraystretch}{1.8}
\begin{array}{l}
\displaystyle T_{s}=RTS+SIFS+\delta+RTC+SIFS+\delta \\
\displaystyle \qquad \  +ATC+SIFS+\delta+CTS+SIFS+\delta \\
\displaystyle \qquad \  +BDATA+SIFS+\delta+BDATA+SIFS+\delta \\
\displaystyle \qquad \  +BACK+SIFS+\delta+BACK+SIFS+\delta \\
\displaystyle T_{c}=RTS+DIFS+\delta
\end{array}
\right.
\end{equation}
Where $\delta$ denotes the propagation delay. Note that $BDATA$  is a little bit longer than $DATA$  due to non-complete overlap of two data frames without synchronization. However, this overhead is negligible and we can approximate $BDATA$ with $DATA$. Also, previous discussion is true for $BACK$.

In a successful two-way relay cooperation, two packets are received by their next two-hop destinations. This is equivalent to four transmissions in the traditional CSMA/CA scheme. Therefore the saturation throughput can be expressed as
\begin{equation}
\phi_{s}=\frac{4P_{succ}E[P]}{X}
\end{equation}
where $E[P]$ represents the average payload in a packet and $P_{succ}$,$X$ are obtained from previous derivations.
\subsection{Saturation Throughput in large-scale network}\label{mhop}
Due to the complexity of large-scale network with hidden node problem, it is difficult to provide an accurate derivation of saturation throughput. Instead, we present an approximate analysis on the throughput performance of this type of network in this subsection. Although several secondary factors are neglected and some assumptions are made, we catch the most important elements that play a significant role in  the performance of the network. Therefore our approximate result can closely reflect the performance of TREAN protocol and provide a guideline for protocol design.

First of all, we determine the transmission failure probability $p_{f}$, defined as the station fails to receive the ACK for its data frame. It should be noticed that the transmission failure not necessarily means the failure of the transmission of payload bits. It is possible that the data frames are received correctly by stations but $BACK$ frames are collided due to hidden stations. In this case, the station increases its backoff stage and takes a random backoff but buffers previous data frame as explained in section (\ref{bmode}).

The transmission failure is caused by two reasons in the network with hidden stations. One of them is the collision of RTS frame. This happens when a station transmit a RTS frame and in the same time slot another station in the interference range of the receiver also begin its transmission. Hence the probability of transmission failure due to RTS collision is given by
\begin{equation}
p_{f1}=1-(1-p_{t})^{n_{i}-1}
\end{equation}
where $n{i}$ denotes the number of stations in the interference range of a station. If the stations are uniformly distributed with density $\lambda$ and the interference range is given by $r_{i}$, the $n_{i}$ can be expressed as
\begin{equation}
n_{i}=r_{i}^2 \pi \lambda
\end{equation}

\begin{figure}[ht]
\centering
\includegraphics[width=0.3\textwidth]{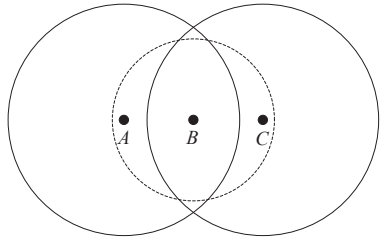}
\caption{\label{srmac} The multiple access phase in the cooperation process. In this phase, station $A$ and station $C$ send message to the relay station B. Solid circles represent the range where corresponding transmitters can be sensed, and the dashed circle denotes the interference range of the receiver}
\end{figure}

\begin{figure}[ht]
\centering
\includegraphics[width=0.3\textwidth]{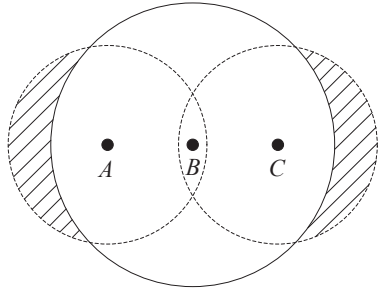}
\caption{\label{srbd} The broadcast phase in the cooperation process. In this phase, the relay station $B$ broadcast information to station $A$ and station $C$. The solid circle represent the range where station B can be sensed, and dashed circles denotes the interference range for receivers. Shadow areas are hidden regions}
\end{figure}

Beside RTS collisions, the transmission failure may happen due to the existence of hidden stations. In the multiple access phases of the two-way relay cooperation process, as shown in Figure \ref{srmac}, stations in the interference range of the relay node can sense at least one ongoing transmission and hence no hidden region exists. In the broadcast phases of cooperation process, corresponding to the transmission of RTC, CTS, BDATA and BACK frames, the transmitter are hidden from two regions indicated by shadow areas in Figure \ref{srmac}.If stations in two shadow regions begin the transmission when the broadcast is underway, the corresponding stations would suffer from collisions. The vulnerable period $T_{v}$ for the collision can be calculated according to different scenarios. As shown in Figure \ref{srbd}, if the broadcast stage follows a transmission from station $A$ (or station $C$), the stations in corresponding hidden region have to wait at least DIFS period after the end of the transmission. Hence the vulnerable period for station $A$ (or station $C$) is given by
\begin{equation}
T_{v}=T_{packet}-(DIFS-SIFS)
\end{equation}
Where $T_{packet}$ is the length of broadcast frame from station B. However, if no transmission from station $A$ (or station $C$) before the broadcast stage, the vulnerable period is simply
\begin{equation}
T_{v}=T_{packet}
\end{equation}
Let $S_{A}$ and $S_{C}$ denote the hidden regions corresponding to station $A$ and station $C$, and $n_{h,S_{i}}$ represent the number of stations in region $S_{i}$ ($i \in \{ A,C \}$). Based on previous discussion, we can determine the vulnerable period $T_{v,j}$ for station $A$ or station $C$ at $j$ ($j \in \{ RTC, CTS, BDATA, BACK\}$) broadcast stage. Then the probability that no collision, due to the transmission initiated by stations in hidden region $S_{i}$, happens at station $i$ in $j$ broadcast stage can be given by
\begin{equation}
(1-p_{t})^{n_{h,S_{i}} \lceil \frac{T_{v,j}}{X} \rceil }
\end{equation}
Where $X$ is generalized time slot as defined in subsection \ref{ssd}.

\begin{figure}[ht]
\centering
\includegraphics[width=0.3\textwidth]{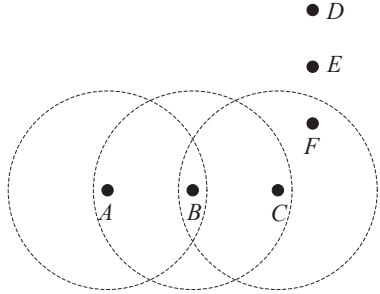}
\caption{\label{pierce} Pierced interference. The station $D$ outside the hidden regions defined previously may initiate a two-way relay cooperation via station $E$ with station $F$ which is in the interference region of station $C$. Hence the transmission of station $F$ may result in the collision on station $C$. We call this type of collisions as pierced interference}
\end{figure}

Actually, the collision may also happens due to the transmission initiated by stations outside the hidden regions defined previously, as shown in Figure \ref{pierce}. However, the requirements for the occurrence of this scenario are demanding: the station $D$ should be not able to sense the transmission of station $A$, station $B$ and station $C$ and initiate the cooperation (i.e. transmit RTS) successfully; In addition, the next two-hop destination station $F$ of the packet from station $D$ should be coincidentally in the interference range of station $A$, station $B$ or station $C$; Also, to respond the cooperation and transmit messages, station $F$ has to receive packets from stations $E$ correctly, i.e. coincidentally avoid from the ongoing transmission from station $A$, station $B$ or station $C$. Therefore the situation does not happen frequently and neglect this case has less impact on the final result.

\begin{figure}[t]
\centering
\includegraphics[width=0.3\textwidth]{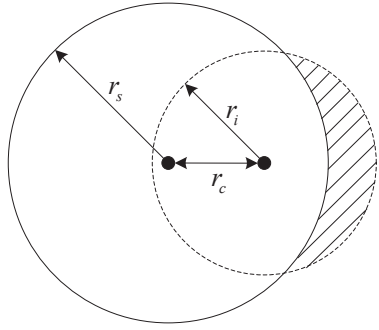}
\caption{\label{hdarea} Hidden region. Let $r_{s}$, $r_{i}$ and $r_{c}$ denote the sensing range, the interference range and the communication range respectively. $n_{h}$ represents the number of stations in shadow region.}
\end{figure}

Based on previous discussion, if we use $n_{h}$ (as shown in Figure \ref{hdarea}) to approximate the number of stations in one hidden region , the probability of transmission failure due to the hidden node problem can be approximated as
\begin{equation}
p_{f2}=1-(1-p_{t})^{n_{1}}
\end{equation}
where
\begin{equation}
n_{1}=2\left(\sum_{j \in \{RTC,CTS,BDATA\}} n_{h}\lceil \frac{T_{v,j}}{X} \rceil\right) +  n_{h}\lceil \frac{T_{v,BACK}}{X} \rceil
\end{equation}
Hence the transmission failure probability is given by
\begin{eqnarray}\label{pf2}
p_{f} & = & 1-(1-p_{f1})(1-p_{f2}) \\
      & = & 1-(1-p_{t})^{n_{i}+n_{1}-1}
\end{eqnarray}

Similar with the derivation of equation (\ref{pc}) in small-scale network cases, the cooperation probability can be expressed as
\begin{equation}\label{pc2}
p_{c}=p_{t}(1-p_{t})^{ni-2+n_{1}}
\end{equation}

To find the value of the generalized time slot $X$, we consider the channel around the station $A$. Let $Z_{A}$ denote the set of stations in the sensing range of station A and $n_{s}$ represent the number of elements in the set $Z_{A}$. Then the probability that the channel is idle can be expressed as
\begin{equation}
P_{idle}=(1-p_{t})^{n_{s}}
\end{equation}

\begin{figure}[th]
\centering
\includegraphics[width=0.2\textwidth]{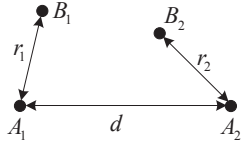}
\caption{\label{cRTS}Concurrent RTS transmission. Consider that $A_{1}$ and $A_{2}$ transmit RTS frame concurrently to $B_{1}$ and $B_{2}$ respectively. If the distance $d$ between two transmitters are great than $(r_{i}+r_{c})$, the distance between $A_{1}$ and $B_{2}$ are great than $d-r_{2}$. Also we know that $r_{2}$ is less than $r_{c}$. Hence the distance between $A_{1}$ and $B_{2}$ are great than $r_{i}$. Therefore there is no interference on station $B_{2}$. This is also true for station $B_{1}$}
\end{figure}

If $A_{i}$ denotes the event that station $i$ transmits a RTS frame successfully, it can be shown that
\begin{equation}
P[A_{i}]=p_{t}(1-p_{t})^{n_{i}-1}
\end{equation}
Then the probability that at least one RTS frame are successfully transmitted is given by
\begin{equation}
P_{RTSsucc}=P\left[\bigcup_{i \in Z_{A}} A_{i}\right]
\end{equation}
We should note that it is possible that in the sensing range of station $A$ two or more RTS can be successfully transmitted concurrently. However, this situation only happens when the transmitters of these RTS are separated enough and hence the collisions are avoided. As explained in Figure \ref{cRTS}, we consider the separation requirement for the concurrent successful RTS transmissions as that the transmitter stands $(r_{i}+r_{c})$ apart with each other.
Let $p_{2c}$ represent the probability that the distance between two randomly chosen points in the disk with radius $r_{s}$ is great than $(r_{i}+r_{c})$. Let $p_{3c}$ represent the probability that the distances between three randomly chosen points in the disk with radius $r_{s}$ are all great than $(r_{i}+r_{c})$. Then it can be shown that
\begin{eqnarray}
P_{RTSsucc} & = & P\left[\bigcup A_{i}\right] \\
            & = & \sum P[A_{i}]-\sum \cap A_{j}]+ \sum P[A_{i} \cap A_{j} \cap A_{k}] \\
            & \approx & \sum P[A_{i}]-\sum_{d(i,j)>r_{c}+r_{i}} P[A_{i} \cap A_{j}]+ \sum_{{d(i,j)>r_{c}+r_{i} \atop d(i,k)>r_{c}+r_{i}}  \atop d(k,j)>r_{c}+r_{i} } P[A_{i} \cap A_{j} \cap A_{k}] \\
            & \approx & n_{s}p_{t}(1-p_{t})^{n_{i}-1}-\binom{n_{s}}{2}p_{2c}p_{t}^{2}(1-p_{t})^{2n_{i}-2}+ \binom{n_{s}}{3}p_{3c}p_{t}^{3}(1-p_{t})^{3n_{i}-3}
\end{eqnarray}
where
\begin{equation}
\left\{
\renewcommand{\arraystretch}{1.8}
\begin{array}{l}
\displaystyle p_{2c}=\int_{d(x,y)>r_{c}+r_{i}} \frac{1}{(r_{s}^2 \pi)^2}\mathrm{d}x\mathrm{d}y \\
\displaystyle p_{3c}=\int_{{d(x,y)>r_{c}+r_{i} \atop d(y,z)>r_{c}+r_{i}}  \atop d(z,x)>r_{c}+r_{i}} \frac{1}{(r_{s}^2 \pi)^3}\mathrm{d}x\mathrm{d}y\mathrm{d}z
\end{array}
\right.
\end{equation}

Then the probability that all RTS frames from stations in $Z_{A}$ collide is given by
\begin{equation}
P_{RTScol}=1-P_{idle}-P_{RTSsucc}
\end{equation}

Let $T_{RTSsucc}$ represents the expectation of channel busy time known that at least a RTS frame from stations in $Z_{A}$ is successfully transmitted. Then the generalized time shot can be expressed as
\begin{equation}\label{X}
X=P_{idle}T_{slot}+P_{RTScol}T_{col}+P_{RTSsucc}E[T_{RTSsucc}]
\end{equation}
As previous discussion, the transmission failure due to the hidden stations may happen at every stage of the two-way relay cooperation after successful RTS transmission. Therefore it is complicated to provide an accurate expression for the expectation of $T_{RTSsucc}$. Hence we take the average of the shortest channel busy time $T_{short}$ and the longest channel busy time $T_{long}$ as an approximation of $E[T_{RTSsucc}]$. Known that RTS frames are successfully transmitted, the shortest channel busy time $T_{short}$ corresponds to the case when RTC frame is collided, while the successful cooperation results in the longest channel busy time $T_{long}$. Hence we have that
\begin{equation}
\left\{
\renewcommand{\arraystretch}{1.8}
\begin{array}{l}
\displaystyle T_{short}=RTS+SIFS+RTC+DIFS \\
\displaystyle T_{long}=T_{s}
\end{array}
\right.
\end{equation}

Based on the equation (\ref{pf2}), (\ref{pc2}), (\ref{X}) and (\ref{pt}), we can solve the transmission failure probability $p_{f}$ with numerical techniques. With this knowledge, the probability  $p_{2s}$ that two data frames are both correctly received by corresponding stations in one round two-way relay cooperation process and the probability $p_{1s}$ that only one data frames are successfully received can be calculated as
\begin{eqnarray}
p_{s2} & = & (1-p_{t})^{n_{i}-1+n_{2}}(1-p_{t})^{n_{h} \lceil \frac{T_{v,BDATA}}{X} \rceil } \\
p_{s1} & = & (1-p_{t})^{n_{i}-1+n_{2}}(2-2(1-p_{t})^{n_{h} \lceil \frac{T_{v,BDATA}}{X} \rceil })
\end{eqnarray}
where
\begin{equation}
n_{2}=2\left(\sum_{j \in \{RTC,CTS\}} n_{h}\lceil \frac{T_{v,j}}{X} \rceil\right) +  n_{h}\lceil \frac{T_{v,BDATA}}{X} \rceil
\end{equation}
Therefore, the saturation throughput for large-scale network is given by
\begin{equation}
\phi_{l}=\frac{Np_{t}(4p_{s2}+2p_{s1})E[P]}{X}
\end{equation}
Where $p_{t}$, $p_{s1}$, $p_{s2}$ and $X$ can be obtained from previous derivations.
\section{Performance Evaluation}
To validate both our physical layer decoding algorithm and MAC protocol design, we perform simulations with MATLAB programme in this section. We first verify the performance of our physical layer decoding algorithm. Then, we evaluate the performance of TREAN protocol in various scenarios.

\subsection{The BER performance of our physical layer decoding algorithm}
\begin{figure}[ht]
\centering
\includegraphics[width=0.65\textwidth]{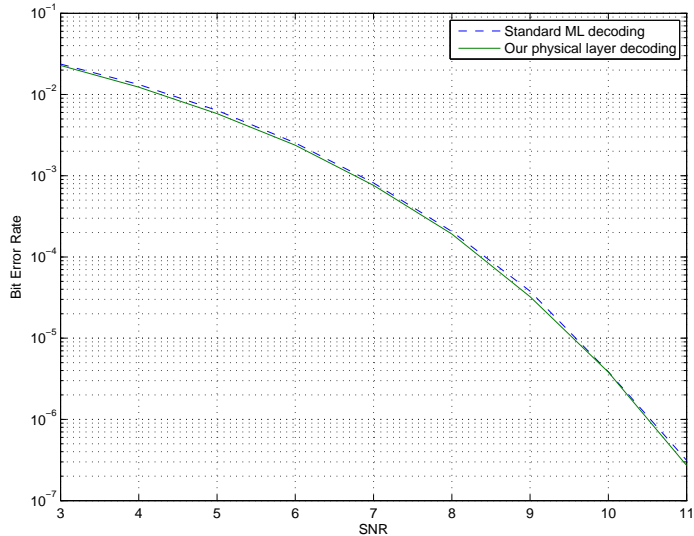}
\caption{\label{ber} Decoding performance in AWGN channel when BPSK modulation is adopted. The dashed line represents the BER performance of standard maximum likelihood decoding when only the desired packet is received. The solid line denotes the BER performance of our physical layer decoding algorithm when the desired packet is received with asynchronous interference from a known packet. }
\end{figure}
Figure \ref{ber} shows that the BER performance comparison between standard ML decoding and our physical layer decoding algorithm in AWGN channel. It can be observed that the decoding performance of our algorithm is quite close to that of standard ML decoding. This demonstrates that our decoding algorithm does not result in any performance loss even the desired packet is received with asynchronous interference from a known packet. Actually, the performance of our decoding algorithm is a little better than that of standard ML decoding. This is due to the diversity gain introduced by oversampling.
\subsection{The performance of TREAN protocol}
In this subsection, we evaluate the performance of TREAN protocol in various settings. The parameters using in the simulation, except the length of control frames in TREAN protocol presented in Table \ref{subtype}, are summarized in Table \ref{par}. All of values given in the table are same with those specified in 802.11 standard \cite{80211} and 802.11a amendment \cite{80211a}. Also, if without explicit indication, we adopt the setting that stations are uniformly distributed in the network in our simulation.
\begin{table}[ht]
\centering
\includegraphics[width=0.35\textwidth]{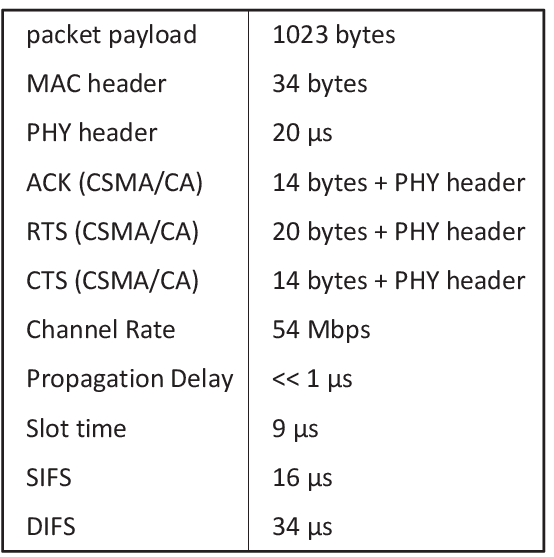}
\caption{\label{par}Parameters using in the simulation}
\end{table}
\subsubsection{Throughput performance of TREAN protocol in small-scale network}
\begin{figure}[ht]
\centering
\includegraphics[width=0.65\textwidth]{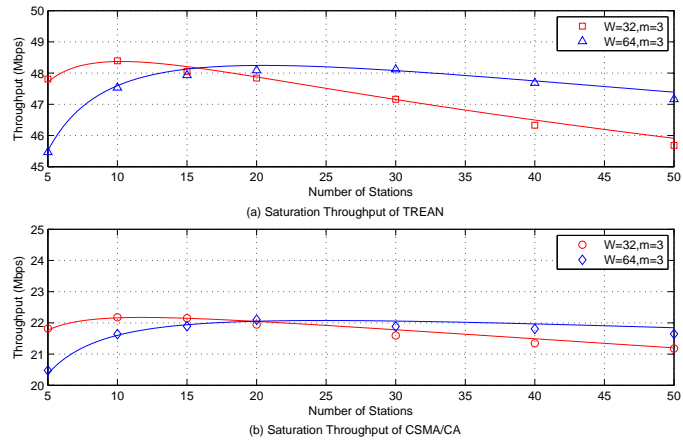}
\caption{\label{sidthr}Throughput performance of TREAN protocol in small-scale network. The symbols denote simulation results which are the average values of 30 repeated experiments, while the solid lines present values calculated from theoretical equations.}
\end{figure}
Figure \ref{sidthr} illustrates the throughput performance of TREAN protocol in small-scale network as the increase of the station number in the network. We can observe that the performance gain of TREAN protocol over traditional CSMA/CA protocol is more than one hundred percent. The main contribution comes from the enhanced spectral efficiency of two-way relay technique. The technique allow two concurrent transmissions instead of one in traditional scheme, and hence double the throughput of the network. Also, the extra benefits of TREAN protocol comes from its more compact transmission manner. In one contention period, four data transmissions (equivalent) are performed in TREAN protocol comparing to only one in CSMA/CA scheme. Hence one data transmission in TREAN protocol partake one fourth contention overhead and one fourth backoff overhead, which is less than one contention overhead and one backoff overhead per data transmission in CSMA/CA protocol.

Also the figure provide the information about the comparison between analytic results and simulation ones. It shows that our theoretical analysis can accurately predict the performance of TREAN protocol, and the error is always less than one percent. This validates our theoretical derivation and hence we can use this analysis tool to further study the performance of TREAN protocol and to optimize parameters such as initial contention window size and the number of stations deployed in the network in practical design.

When perfect scheduling is adopted, the saturation throughput of the network with two-way relaying technique can approach 86\% of the ideal rate 108Mbps. The fourteen percent performance loss is introduced by MAC layer header and PHY layer header added to payload bits. With random access MAC protocol TREAN, the saturation throughput is about 45\% of the ideal rate. It seems that the application of two-way relaying in a scalable manner is at the cost of the performance degradation. However, the same problem exists in traditional CSMA/CA protocol. The saturation throughput of the network with 802.11 MAC layer is only 41\% of the ideal rate 54Mbps comparing to 86\% in the perfect scheduling case. Therefore the performance degradation is due to the inefficiency of CSMA/CA protocol itself (note that TREAN protocol is based on the CSMA/CA protocol). However several research works have been done to address this issue. In \cite{wifinano}, the efficiency of CSMA/CA protocol is boosted to 80\% when using the rate 54Mbps. If we incorporate those sophisticated techniques in \cite{wifinano} in TREAN protocol, the efficiency of our protocol would approach the perfect scheduling solution. However, this is out of the scope of this paper.
\subsubsection{Throughput performance of TREAN protocol in large-scale network}
\begin{figure}[ht]
\centering
\includegraphics[width=0.58\textwidth]{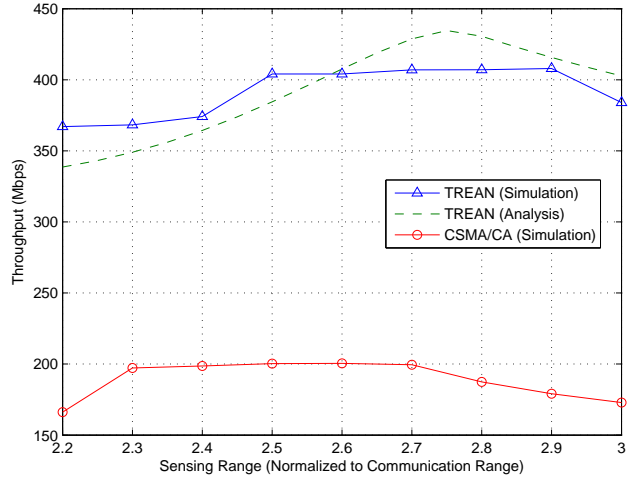}
\caption{\label{multithr}Throughput performance of TREAN protocol in large-scale network. The symbols denote simulation results which are the average values of 15 repeated experiments, while the dashed line present values calculated from analytic equations. }
\end{figure}
Figure \ref{multithr} shows the throughput performance of TREAN protocol in large-scale network as the increase of the sensing range. The results are obtained in the setting that the network contains three hundred stations and have size $10 \times 10$ if the length of communication radius is set to 1. Also, the interference range is given as 1.78 according to \cite{ri}.

It can be observed that the throughput gain of TREAN protocol over CSMA/CA scheme is about one hundred percent. This is partially due to the more compact special reuse of TREAN protocol, i.e. two-way relay allows two concurrent transmissions which is quite close to each other. However this is impossible in CSMA/CA protocol. Hence, on average, more concurrent transmissions can coexist under TREAN protocol than under CSMA/CA protocol. Also, the reduced contention and backoff overhead per data transmission as mentioned previously contribute to the enhancement of the performance. This contribution is more significant in large-scale network comparing to that in small-scale network. This is because that backoff time slots of different stations are not synchronized with each other due to the fact that different stations have different sensing regions in large-scale network and hence have different sensing results. Therefore stations access the channel not in a slotted manner in large-scale network, and this results in more frequent RTS collisions and hence more contention overhead.

In addition, similar with the CSMA/CA case, the throughput of the network under TREAN protocol first goes up and then decline with the increase of the sensing range. The increment of the throughput as the increase of sensing range is due to the notable decrease of collisions caused by hidden stations, while the decline of the performance when the sensing range further increases is because of more conservative spatial reuse. Also, we note that the optimal sensing range in TREAN protocol is larger than that in CSMA protocol. This is because the structure unit of physical layer under TREAN protocol is larger and more complicated than that of traditional physical layer and hence suffer more serious hidden node issue. Therefore we need larger sensing range to diminish the effect due to hidden stations in TREAN protocol.

Figure \ref{multithr} also show the comparison between theoretic throughput and simulation results. It can be observed that our theoretic results can closely reflect the variation trend of throughput with the increase of the sensing range and the error comparing to the simulation results is always less than five percent. Hence our approximated derivation on the saturation throughput of large-scale network is enough to give a preliminary estimation on the performance of TREAN protocol and can provide sufficient insights and guidelines for the design in realistic systems.
\subsubsection{Extended Modes of TREAN protocol}
\begin{figure}[ht]
\centering
\includegraphics[width=0.58\textwidth]{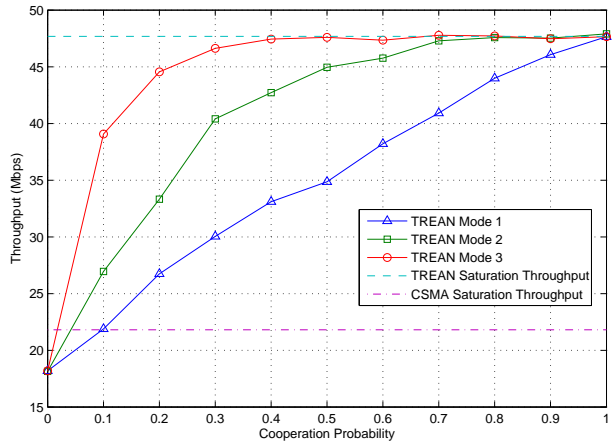}
\caption{\label{sc1234}Throughput performance of different modes of TREAN protocol}
\end{figure}
Figure \ref{sc1234} shows the throughput of the network when running different modes of TREAN protocol. In this simulation, the network is a small-scale one with forty stations in it. Also, the extended modes is implemented with two EA fields in an extended RTC frame, and the management overhead is not taken into consideration. We can find that the extend modes have much better performance than the basic mode of TREAN protocol in low ATC-probability region as expected. In addition, when the ATC-probability is zero, different modes of TREAN protocol have same performance and are about twenty percent worse than CSMA/CA protocol. This is partially due to the larger overhead of control frames in TREAN protocol. Also, in zero ATC-probability scenario, only one-way relay transmission is performed. This scheme is even more vulnerable to the hidden node issue and hence results in the performance degradation of TREAN protocol.
\subsubsection{CPP frame in TREAN protocol}
\begin{figure}[ht]
\centering
\includegraphics[width=0.3\textwidth]{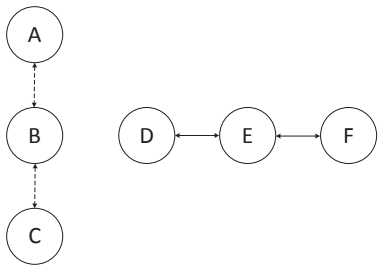}
\caption{\label{simufair} A network topology potential with the fairness problem. There are six station $A$, $B$, $C$, $D$, $E$, $F$ in the network. The distances between $AB$, $BC$, $DE$, $EF$ are all 0.8 and the distance between $BE$ is 1.2 if the communication range is set to 1. We assume that station $A$ and station $C$ have data frames to exchange, so do station $D$ and station $F$. }
\end{figure}

\begin{figure}[ht]
\centering
\includegraphics[width=0.53\textwidth]{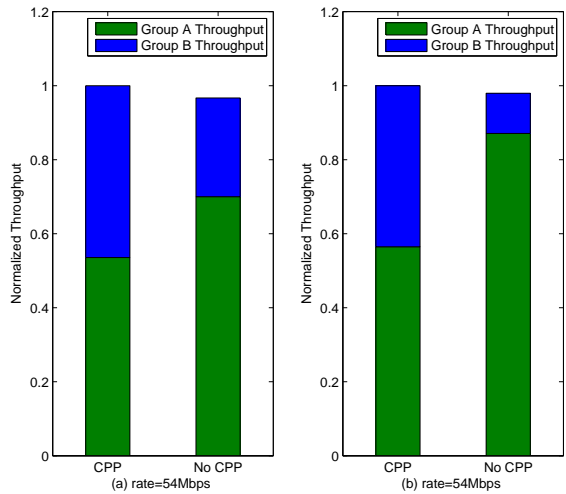}
\caption{\label{cppfair} The comparison between the protocol with CPP and without CPP}
\end{figure}
Figure \ref{cppfair} reveals the significance of the CPP frame for protecting captured channel. The simulation is run in a network potential with the fairness problem as shown in Figure \ref{simufair}. The interference range is set 1.78 as the previous simulation and the sensing range is equal to 2.9 which is optimal value for TREAN protocol. We can observe that when the CPP frame is present, the throughput of group 1 is comparable to that of group 2. However if we disable the CPP frame in the protocol, the throughput of group 1 degrades dramatically but the total throughput keeps unchanged. This indicates that the channel got by station $D$ is recaptured by station $A$ or station $C$ without the protection of the CPP frame. Also, we note that the situation is more worse for low communication rate scenario. This is because that the lower rate results in longer control frames and hence increases the vulnerable period when the channel may be recaptured.
\subsubsection{ACK loss rate}
\begin{table}[ht]
\centering
\begin{tabular}{|c|c|c|c|c|c|}
\hline
\textbf{sensing range} & 2.2 & 2.4 & 2.6 & 2.8 & 3.0 \\ \hline
\textbf{ACK loss rate} & 15.12\% & 12.98\% & 6.39\% & 3.82\% & 1.59\% \\ \hline
\end{tabular}
\caption{\label{ackloss}ACK loss rate}
\end{table}
Table \ref{ackloss} summarize the ACK loss rate for different sensing ranges in large-scale network with TREAN protocol. It can found that the ACK loss rate is reduced as the increase of sensing range. This is due to the fact hidden node issue which results in the loss of ACK is mitigated when the sensing range increases.

With our ACK scheme, the correctly received data frame is not acknowledged only when all three ACK frames with the ID of this data frame are lost. The probability that this situation happens is much lower than one percent even the sensing range is only 2.2 times as the communication range, and hence the negative impact of ACK loss on the performance is almost negligible.
\section{Conclusion}
In this paper, we propose a complete, practical and high-performance solution for applying the two way relaying in general ad hoc networks. The solution includes a new physical layer scheme for two way relaying and a random access MAC protocol TREAN. The new physical layer scheme does not require any synchronization and is feasible for any linear modulation schemes. This lays a solid foundation for the design of a wide applicable random access MAC protocol. On the top of this physical layer scheme, a 802.11-like protocol TREAN is proposed. TREAN protocol support two-way relaying technique to boost the performance of the network but avoid complicated scheduling and optimization. Hence it can be adopted in large scale network. Furthermore, to guarantee the performance gain even when bi-directional data flows are absent, extended modes of TREAN protocol are suggested. Simulation results and theoretical analysis both shows that our integrated solution can provide remarkable improvement on the performance comparing to traditional CSMA/CA protocol.

The implementation of our solution requires the change on physical chip due to the redesign of the physical layer. However, the operations in our physical layer schemes only involve common ones such as correlation, matrix computation and interpolation. All of these can be implemented with standard DSP module. In addition, TREAN protocol is based on standard CSMA/CA protocol and hence can be obtained from the modification of existing system. The detailed implementation of our solution is left as our future work.

\end{document}